\documentclass[conference]{IEEEtran}
\IEEEoverridecommandlockouts

\usepackage{cite}
\usepackage{amsmath,amssymb,amsfonts,amsthm}
\usepackage{algorithmic}
\usepackage{graphicx}
\usepackage{textcomp}
\usepackage{xcolor}
\usepackage{booktabs}
\usepackage[skip=5pt]{caption}
\usepackage{subcaption}
\usepackage{multirow}
\usepackage{diagbox}
\usepackage{url}
\usepackage{dsfont}
\usepackage{enumitem}
\usepackage{balance}
\usepackage{tabularx}
\usepackage{svg}
\usepackage{float}
\usepackage{weiwAlgorithm}
\usepackage{array}
\usepackage{booktabs}

\usepackage[colorlinks,citecolor=black,linkcolor=black,urlcolor=black]{hyperref}

\usepackage{diagbox}

\def\BibTeX{{\rm B\kern-.05em{\sc i\kern-.025em b}\kern-.08em
    T\kern-.1667em\lower.7ex\hbox{E}\kern-.125emX}}

\theoremstyle{definition}
\newtheorem{definition}{Definition}

\newtheorem{example}{Example}

\theoremstyle{plain}
\newtheorem{theorem}{Theorem}
\newtheorem{lemma}{Lemma}
\newtheorem{observation}{Observation}
\newtheorem{corollary}{Corollary}

\newcommand{\myparagraph}[1]{\noindent \textbf{#1.}}
\newcommand{\myparagraphexp}[1]{\noindent \textbf{#1.}}

\begin{document}

\title{HL-index: Fast Reachability Query in Hypergraphs}



\author{
Peiting Xie$^{\dag}$, Xiangjun Zai$^{\dag}$, Yanping Wu$^{\ddag}$, Xiaoyang Wang$^{\dag}$, Wenjie Zhang$^{\dag}$, Lu Qin$^{\ddag}$ 
\vspace{1mm}\\
\fontsize{9}{9}\selectfont\itshape
$^{\dag}$\textit{The University of New South Wales}, Australia 
$^{\ddag}$\textit{University of Technology Sydney}, Australia 
\vspace{1mm}\\
\fontsize{9}{9}\selectfont\ttfamily\upshape
\{peiting.xie,xiangjun.zai,xiaoyang.wang1,wenjie.zhang\}@unsw.edu.au \\
yanping.wu@student.uts.edu.au
lu.qin@uts.edu.au
}


\maketitle







\begin{abstract}

Reachability in hypergraphs is essential for modeling complex groupwise interactions in real-world applications such as co-authorship, social network, and biological analysis, where relationships go beyond pairwise interactions. In this paper, we introduce the notion of $\boldsymbol{s}$-reachability, where two vertices are $\boldsymbol{s}$-reachable if there exists a sequence of hyperedges (i.e., a walk) connecting them, such that each pair of consecutive hyperedges shares at least $\boldsymbol{s}$ vertices. Moreover, we define the max-reachability query as a generalized form of the $\boldsymbol{s}$-reachability problem, which aims to find the largest value of $\boldsymbol{s}$ that allows one vertex to reach another. To answer max-reachability queries in hypergraphs, we first analyze limitations of the existing vertex-to-vertex and hyperedge-to-hyperedge indexing techniques. We then introduce the HL-index, a compact vertex-to-hyperedge index tailored for the max-reachability problem. To both efficiently and effectively construct a minimal HL-index, we develop a fast covering relationship detection method to eliminate fruitless hypergraph traversals during index construction. A lightweight neighbor-index is further proposed to avoid repeatedly exploring neighbor relationships in hypergraphs and hence accelerate the construction. Extensive experiments on 20 datasets demonstrate the efficiency and scalability of our approach.

\end{abstract}


\maketitle




\section{Introduction}\label{sec:intro}
\begin{figure}[t]
  \centering
  \begin{minipage}{0.8\linewidth} 
    \centering
    \begin{subfigure}[b]{0.49\linewidth}
      \centering
      \includegraphics[width=\linewidth]{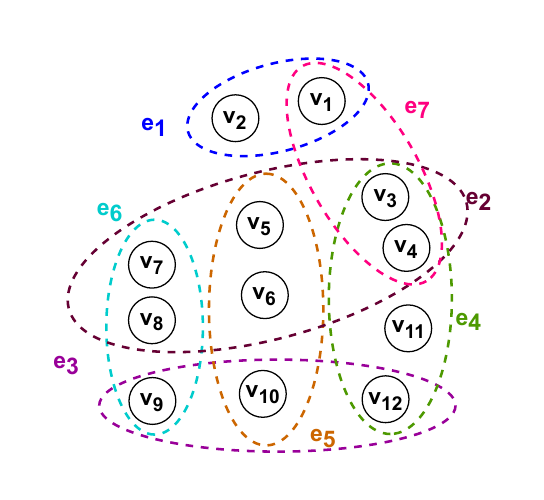}
      \caption{Hypergraph $\mathcal{H}$}
    \end{subfigure}
    \hfill
    \begin{subfigure}[b]{0.49\linewidth}
      \centering
      \includegraphics[width=\linewidth]{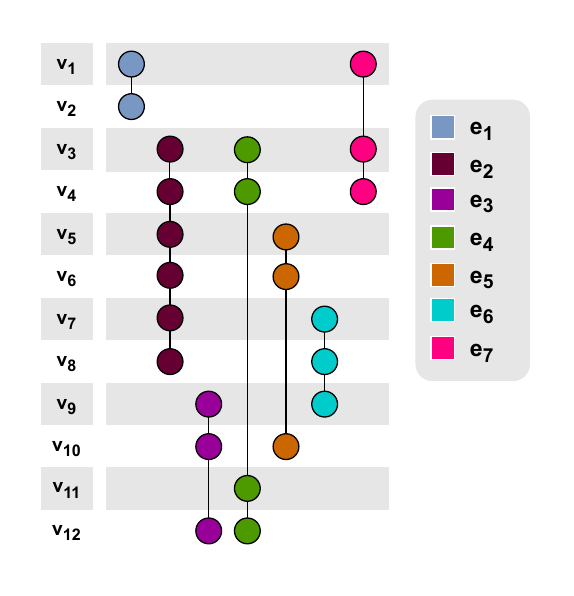}
      \caption{Vertices in hyperedge}
    \end{subfigure}
    \caption{An example of hypergraph $\mathcal{H}$}
    \label{fig:ex}
  \end{minipage}
\end{figure}

Reachability is a fundamental problem in graph analysis that asks whether a path exists between two vertices. 
It has been extensively studied over the past decades~\cite{agrawal1989efficient,cohen2003reachability, su2016reachability, zhu2014reachability}. 
However, existing works mainly focus on general graphs, where each edge connects exactly two vertices,
ignoring higher-order relationships that arise from group-based interactions.
In real-world applications, many systems involve complex group relationships, where multiple entities interact simultaneously~\cite{cui2013online, luo2023toward, feng2021hypergraph, yook2004functional, hartwell1999molecular}. 
For example, Figure~\ref{fig:ex}(a) shows a co-authorship network, 
where each vertex represents an author.
The vertices within each dashed oval show a group of individuals working together on the same task, 
such as co-authoring a paper.
In biological networks, proteins often work together to achieve specific biological functions.
The above scenarios can be modeled as a hypergraph $\mathcal{H}=(\mathcal{V},\mathcal{E})$, where
$\mathcal{V}$ denotes the set of vertices and 
each hyperedge $e\in \mathcal{E}$ is a non-empty subset of $\mathcal{V}$.

As a natural data representation to exhibit multi-ways relationships beyond simple binary ones, hypergraphs have gained significant research attention and have been studied in different areas,  
such as recommendation \cite{bu2010music, zheng2018novel, zhu2016heterogeneous}, social analysis \cite{tan2014mapping, yang2019revisiting}, bioinformatics \cite{hwang2008learning} and e-commence~\cite{li2018tail}.  
To model complex polyadic interactions in hypergraphs, we adopt the notion of \textit{s}-walks \cite{preti2024hyper, aksoy2020hypernetwork, pan2025hitec}. 
A \textit{s}-walk in a hypergraph is a sequence of hyperedges, where every two consecutive hyperedges share at least \textit{s} vertices. 
A larger value of the overlap \textit{s} reflects a stricter notion of adjacency, thereby indicating stronger underlying relationships among groups \cite{preti2024hyper}. 
\cite{aksoy2020hypernetwork} demonstrates the power of 
\textit{s}-walks in uncovering meaningful and interpretable insights within higher-order data, which are often overlooked by traditional pairwise graphs. 

Motivated above, in this paper, we introduce a new reachability query, named \textit{$s$-reachability}, in hypergraphs. 
Specifically, given a hypergraph $\mathcal{H=(V,E)}$, two vertices $u$, $v \in$ $\mathcal{V}$ and a positive integer $s$, 
we say $u$ can \textit{s}-reach $v$, denoted by $u \stackrel{s}{\leadsto} v$,
if there exists a $s$-walk $\{e_u, e_1, ..., e_v\}$ that satisfies $u \in e_u$ and $v\in e_v$.
To further measure how strongly two entities are connected, for two query vertices $u,v$, we propose the \textit{max-reachability} problem, which aims to compute the largest value of $s'$ such that $u \stackrel{s'}{\leadsto} v$ holds.
Note that, max-reachability can be regarded as a generalized model of $s$-reachability, since once the maximum overlap value is obtained, all $s$-reachability queries can be answered directly by comparing $s$ with this value.
Therefore, in this paper, we mainly focus on the max-reachability problem.
Below is an example of max-reachability.

\begin{example}\label{example:1}
Reconsider the hypergraph $\mathcal{H}$ shown in Figure~\ref{fig:ex}(a). 
The vertices within each hyperedge in $\mathcal{H}$ are explicitly detailed in Figure~\ref{fig:ex}(b).
Suppose the query vertices are $v_5$ and $v_{9}$, and the value of $s$ is 2. 
There exists a $2$-walk $\{e_2, e_6\}$ between $v_5$ and $v_{9}$, where $v_5 \in e_2$ and $v_9\in e_6$.
Therefore, 
$v_5$ can 2-reach $v_9$, i.e., $v_5 \stackrel{2}{\leadsto} v_9$.
When applying $s = 3$, 
no 3-walk between $v_5$ and $v_9$ exists.
Consequently, the max-reachability between these two vertices is 2.
\end{example}



The proposed model can be adopted in many real-world applications, and we list some of them as follows.

\begin{itemize}[leftmargin=*]
    \item  In the field of internet communication,
    the multi-hop wireless network can be represented as a hypergraph~\cite{DBLP:journals/ton/GaoZRSRB15}, where vertices represent devices and hyperedges represent a set of directly reachable devices.
    The hypergraph model is suitable for opportunistic routing schemes, such as ExOR, GeRaF, and MORE~\cite{DBLP:conf/sigcomm/BiswasM05}. In these schemes, any device that receives a packet is eligible to forward it, with receiving devices typically coordinating through a protocol to decide which device will forward the packet. 
    By assessing the $s$-reachability between two devices on the hypergraph, it is possible to ensure a more robust communication path, avoiding weak connections and improving network reliability.
    Besides, identifying the maximum overlap value among all paths between two devices can help optimize resource allocation by focusing on the most cohesive routes, further enhancing the efficiency and scalability of the network.
    \item Hypergraphs can be used to model 
    biological networks, where vertices represent proteins and hyperedges represent the group of proteins working together~\cite{klamt2009hypergraphs}. Such group interactions between proteins achieve some well-defined biological function, i.e., functional modular~\cite{yook2004functional, hartwell1999molecular}. 
    This hypergraph-based representation allows us to uncover potential associations between proteins across multiple modules.
    By determining the $s$-reachability between two proteins, we can evaluate the robustness of the functional connectivity between them.
    Furthermore, identifying the max-reachability value can 
    help prioritize the strongest functional interactions, providing insights into critical biological pathways.
    These studies can be valuable in disease predictions, since they enable researchers to identify potential weak links in biological processes that may lead to cascade effects~\cite{albert2000error}.
    \item 
    Hypergraphs can also be used to analyze disease transmission networks \cite{higham2021epidemics}, where vertices represent individuals and hyperedges represent groups of people co-located within a given time window. By constructing a hypergraph over the period between the earliest and latest reported infections, we can evaluate the max-reachability between patients to analyze potential transmission relations, and further estimate the likelihood of healthy individuals being infected. A larger max-reachability value indicates stronger potential contagion connections, suggesting that patients are more likely to share an underlying transmission chain or infection relation. This enables researchers to identify high-risk groups and better understand the spread dynamics of diseases.
\end{itemize}

\noindent \textbf{Challenges and our approaches}.
To our best knowledge, we are the first to propose and investigate the max-reachability problem in hypergraphs. 
To answer max-reachability queries, we first design an online solution, which performs a bidirectional priority-based search from query vertices $u,v$ and always traverses the hyperedge that forms the walk with the current highest overlap.
During traversal, 
it records the highest overlap for every visited hyperedge from $u/v$ and updates the answer when a hyperedge is visited from both sides. While effective, this online approach is time-consuming, as it may need to traverse the whole hypergraph for a single query. 

To address the inefficiency in the online method, a common approach is to design an efficient indexing mechanism to perform rapid query processing. 
In the literature, various index structures are proposed to improve the efficiency of reachability queries, where the 2-hop labeling framework \cite{cohen2003reachability} has been widely adopted. 
Previous studies leverage the 2-hop labeling framework to maintain the vertex-to-vertex relationships
and answer reachability queries accordingly, e.g.,~\cite{wen2022span, chen2021efficiently}. 

However, it is non-trivial to apply this vertex-to-vertex based scheme to our problem,
since this strategy may lead to incorrect results. 
Alternatively, a natural idea is to maintain the reachability information between hyperedges to ensure correctness.
Unfortunately, this approach is impractical, since each vertex may belong to multiple hyperedges, and a prohibitive query cost will be incurred.
Moreover, it is infeasible to maintain neighbor information globally due to the impractical space consumption, while computing neighbors in runtime for hypergraph traversal is severely time inefficient.
To efficiently answer max-reachability queries, 
inspired by the 2-hop labeling framework, we propose a novel vertex-to-hyperedge based index, named \underline{H}ypergraph \underline{L}abeling index (HL-index). Specifically, for each vertex $u$, we maintain a label set $\mathcal{L}(u)$. 
Each element in $\mathcal{L}(u)$ is a pair $(e, s)$, which indicates $u$ can $s$-reach hyperedge $e$.
Given two query vertices $u, v$ and a positive integer $s$, we deduce that $u$ can $s$-reach $v$, if there exists a common hyperedge $e$ that $(e, s_u) \in \mathcal{L}(u)$ and $(e, s_v) \in \mathcal{L}(v)$, where $s_u, s_v \ge s$. The max-reachability between $u$ and $v$ can be derived by iterating through every element in $\mathcal{L}(u)$ and $\mathcal{L}(v)$ and recording the highest $s$ value obtained. 
Note that, efficiently constructing a minimal HL-index is non-trivial. 
Naively extending the previous 2-hop construction methods cannot effectively determine the necessity of entries during index construction and hence violates the minimality of the index. 
Moreover, due to the high expense of hyperedge neighbor computation, traversal in hypergraphs is much more costly than in {pairwise} graphs.
To tackle the above issues, in this paper, we study the dominance relationship between vertices and hyperedges. By leveraging the characteristics of undirected graphs, we explore an efficient covering relationship detection method to prune redundant hypergraph traversals. We further accelerate the index construction process by developing a novel data structure called neighbor-index.
This lightweight neighbor-index dynamically maintains all useful neighbor information during the hypergraph traversal, enabling orders of magnitude smaller memory consumption compared with pre-computing and storing neighbor information for all hyperedges. Moreover, an innovative approach is designed to ensure the minimality of our HL-index. 
The contributions of this paper are summarized as follows. 
\begin{itemize}[leftmargin=*]
    \item \underline{Novel reachability model in hypergraphs.} To capture the properties of groupwise connections between vertices, we introduce a new hypergraph reachability model based on the concept of $s$-walk and conduct the first research to investigate 
    the max-reachability problem in hypergraphs. 
    \item \underline{Efficient index-based solution.} To address this challenging problem, we develop a novel labeling framework called HL-index that maintains the relationships between vertices and hyperedges, which prevents the exhaustive computations between multiple hyperedges. We also propose two effective optimizations to speed up the index construction, i.e., fast covering relationship detection and a lightweight neighbor-index structure to support efficient hypergraph traversals. 
    \item \underline{Minimality of HL-index.} 
    To ensure the minimality of our HL-index,
    we propose an efficient strategy of examining whether a label in our HL-index is redundant, instead of exhaustively invoking the correctness checking for every pair of vertices after removing the corresponding label.
    \item \underline{Extensive performance studies on real-world datasets.} 
    We conduct comprehensive experiments on 20 datasets.
    The result demonstrates that our HL-index outperforms all baselines in terms of query time efficiency and can achieve up to 3 orders of magnitude speedup. A case study in epidemic analysis further demonstrates its utility in risk quantification.
\end{itemize}

\textit{Note that, due to the limited space, all the omitted proofs, partial implementation details, and additional experimental results can be found in the online Appendix~\cite{appendix}.}




\section{Preliminary} \label{sec:2}



Let $\mathcal{H}=(\mathcal{V}, \mathcal{E})$ denote a hypergraph, where $\mathcal{V}$ and $\mathcal{E}$ represent the set of vertices and hyperedges, respectively. 
The number of vertices and hyperedges are denoted by $n = \left|\mathcal{V}\right|$ and $m = \left|\mathcal{E}\right|$, respectively. 
Each hyperedge $e$ is a non-empty subset of $\mathcal{V}$, 
and the number of vertices in $e$ is marked as $\left|e\right|$.
We use 
$E(u)$ to represent the set of hyperedges in $\mathcal{E}$ that contains the vertex $u$, i.e., $E(u) = \{e_i | u \in e_i \land e_i \in \mathcal{E}\}$. 
The degree of vertex $u$ is denoted by $|E(u)|$.
Given two hyperedges $e_i$, $e_j \in \mathcal{E}$, we say $e_i$ and $e_j$ are neighbors if  $e_i \cap e_j \ne \emptyset$.
For a hyperedge $e$, we use $N(e)$ to denote its neighbor set. 
The maximum and average numbers of hyperedges incident to a vertex are denoted by $\eta_{max}$ and $\eta_{avg}$, respectively. Table \ref{tab:notation} lists the notations frequently used in this paper.

\begin{table}[t]
\caption{Frequently used notations}
\label{tab:notation}
\footnotesize
\centering
\resizebox{0.98\linewidth}{!}{%
\begin{tabular}{lcccc}
\toprule
\textbf{Notation} & \textbf{Description} \\
\midrule
$\mathcal{H} = \mathcal{(V, E)}$ & the hypergraph \\
$E(u)$ & set of hyperedges in $\mathcal{E}$ that contains $u$  \\
$N(e)$ & set of hyperedges in $\mathcal{E}$ that are neighbors of $e$  \\
$MR(u,v)$ & max-reachability query between $u$ and $v$ \\
$\eta_{max}$, $\eta_{avg}$ & maximum/average node degree  \\
$d$, $\delta$ &
$\max_{v \in \mathcal{V}}\left|E(v)\right|$, $\max_{e \in \mathcal{E}}\left|e\right|$\\
$W(e_i,e_j)$ & walk from $e_j$ to $e_j$ \\
$\mathcal{W}(e_s, e_t)$ & set of all walks from $e_s$ to $e_t$  \\
$OD(e_i, e_j)$ &  overlapping degree of hyperedges  \\
$WOD(W)$ & walk overlap degree of $W$ \\
$MCD(e)$ & maximum covering degree of $e$  \\
\bottomrule
\end{tabular}
}

\end{table}






%
\begin{definition}[Overlapping degree]
    Given a hypergraph $\mathcal{H}=(\mathcal{V}, \mathcal{E})$ and two hyperedges $e_i, e_j \in \mathcal{E}$, the \underline{o}verlapping \underline{d}egree between $e_i$ and $e_j$, denoted by $OD(e_i, e_j)$, is the number of common vertices in $e_i$ and $e_j$, i.e., $OD(e_i, e_j) = \left|e_i \cap e_j \right|$.
\end{definition}

Given a hypergraph $\mathcal{H} = (\mathcal{V, E})$ and two hyperedges $e_i, e_j \in \mathcal{E}$, a \textit{walk} \cite{preti2024hyper} from $e_i$ to $e_j$, denoted by $W(e_i,e_j)$, is a sequence of hyperedges $\{e_0, e_1,\dots,e_{l-1}, e_l\}$, such that $i)$
$e_i = e_0$, $e_j = e_l$, and $ii)$ $\forall_{1\le i\le l}, OD(e_{i-1}, e_i) \ge 1$.
We use $\mathcal{W}(e_s, e_t)$ to denote the set of all walks from $e_s$ to $e_t$. When the context is clear, we omit the source and destination for those walks, where $W(e_i, e_j)$ will be simplified as $W$. We further use the symbol $\oplus$ to denote the walk concatenation, where $W_i \oplus W_j$ refers to concatenating $W_j$ to the end of $W_i$. Given the concept of walk, we have the following definitions.


\begin{definition}[Walk overlapping degree]
    Given a walk $W = \{e_0, e_1, ...,e_{l-1}, e_l\}$, the \underline{w}alk \underline{o}verlapping \underline{d}egree of $W$, denoted by $WOD(W)$, is the minimum overlapping degree of any two consecutive hyperedges in $W$, i.e., $WOD(W) = \min_{1 \le i \le l}OD(e_{i-1}, e_{i})$.
\end{definition}

\begin{example}
As shown in Figure \ref{fig:ex}(a), we have a walk $W = \{e_2, e_5, e_3\}$ with $WOD(W)$ of 1, since $e_2$ and $e_5$ share the vertices $v_5$ and $v_6$, while only one vertex $v_{10}$ exists in both $e_5$ and $e_3$. Therefore, we have $WOD(W) =  {\min(2,1)} = 1$.
\end{example}



A walk is an $s$-walk iff for every pair of consecutive hyperedges shares at least $s$ common vertices. If the walk contains only one hyperedge $e$, its WOD is defined as $\left|e\right|$.
Given two hyperedges $e_i$, $e_j$, if there exists a $s$-walk $W(e_i, e_j)$ s.t. $WOD(W(e_i, e_j)) \ge s$, we use $e_i \stackrel{s}{\leadsto} e_j$ to denote such reachability relationship between $e_i$, $e_j$.
On this basis, we define the $s$-reachability between vertices as follows.

\begin{definition}[$s$-reachability]\label{th:rule2} 
    Given a hypergraph $\mathcal{H}=(\mathcal{V}, \mathcal{E})$, two vertices $u$, $v\in \mathcal{V}$ and a positive integer $s$, we say $u$ can $s$-reach $v$, denoted by $u \stackrel{s}\leadsto v$, iff there is a walk $W(e_i, e_j)$ s.t. $i)$ $u\in e_i$, $v\in e_j$ and $ii)$ $WOD(W(e_i, e_j)) \ge s$.
\end{definition}

It is easy to verify that $u \stackrel{s}\leadsto v$ equals $v \stackrel{s}\leadsto u$.
Also, a vertex $u$ can $s$-reach a hyperedge $e$, 
if there exists a hyperedge $e'$ s.t. $u \in e' \land$ $e' \stackrel{s}{\leadsto} e$, which is denoted by $u \stackrel{s}\leadsto e$.



\begin{corollary}\label{th:corollary}
    Given two hyperedges $e_i, e_j$, if $e_i \stackrel{s}\leadsto e_j$, then $\forall u \in e_i, \forall v \in e_j$, we have $u \stackrel{s}{\leadsto} v, u \stackrel{s}{\leadsto} e_j$ and $e_i \stackrel{s}{\leadsto} v$. If $u$ and $v$ belong to the same hyperedge $e$, we have $u \stackrel{|e|}\leadsto v$.
\end{corollary}






\myparagraph{Problem 1} Given a hypergraph $\mathcal{H}=(\mathcal{V}, \mathcal{E})$, two vertices $u, v \in \mathcal{V}$ and a positive integer $s$, we want to query if $u \stackrel{s}{\leadsto} v$.

\begin{example}
Given the hypergraph in Figure \ref{fig:ex}(a), 
querying whether 
$v_{1} \stackrel{2}\leadsto v_{10}$ returns true, since $v_1$ can reach $v_{10}$ by the walk $W =\{e_7, e_2, e_5\}$ with $WOD(W) = 2$.

\end{example}

The value of $s$ can be used to model the strength of association between two vertices in the hypergraph. 
To better illustrate the $s$-reachability relationship between vertices, we further introduce a generalized model as follows.

\begin{definition} [Max-reachability] \label{th:def4}
    Given a hypergraph $\mathcal{H}=(\mathcal{V}, \mathcal{E})$, $u, v \in \mathcal{V}$, 
    let $e_{pair}(u,v) = \{(e_u, e_v)| u\in e_u \land v\in e_v\}$ be the set of all hyperedges pairs containing $u$ and $v$. Then,     
    the \underline{m}ax-\underline{r}eachability $MR(u, v)$ between $u$ and $v$ is defined as 
    $MR(u, v) = \max\{WOD(W)| W \in \mathcal{W}(e_w) \land e_w \in e_{pair}\}$.
    
\end{definition}


\myparagraph{Problem 2}  Given the hypergraph $\mathcal{H=(V, E)}$, two vertices $u, v \in \mathcal{V}$, we aim to compute $MR(u, v)$. 

\vspace{1mm}
\textit{Note that, 
Problem 2 is a generalized version of Problem 1, since we have $u \stackrel{s}{\leadsto} v$ iff $MR(u,v) \ge s$. 
Therefore, 
in this paper, we mainly focus on Problem 2,
whose solution can be extended to solve Problem 1 as discussed in Section~\ref{sec:5}.}


\section{Online Solution} \label{sec:sec3}



We first present an online
algorithm for the max-reachability problem.
Generally, it performs a bidirectional priority-based search, starting from query vertices $u$ and $v$, respectively.
These two priority queues store explored hyperedges, allowing us to always traverse the hyperedge that forms the walk with the current highest WOD value.
The result is updated once a hyperedge is visited from both sides. Exploration for a walk is terminated if its WOD is no larger than the result.

\begin{algorithm}[t]
	{
		\SetVline
		\footnotesize
		\caption{Online Search Algorithm}
		\label{alg:BASELINE}
		\Input{A hypergraph $\mathcal{H(V, E)}$, source vertex $u$, target vertex $v$}
		\Output{The result of $MR(u, v)$}
        \State{ $visit_{in}[e] \gets -1$, $visit_{out}[e] \gets -1$ \textbf{for each} $e\in \mathcal{E}$}
        \State{$Q_{in}\gets \emptyset$, $Q_{out} \gets \emptyset$, $result \gets 0$}
        
        \State {$Q_{out} \leftarrow Q_{out} \cup (e, \left|e\right|)$ \textbf{for each} $e \in E(u)$}

       \State {$Q_{in} \leftarrow Q_{in} \cup (e', \left|e'\right|)$ \textbf{for each} $e' \in E(v)$}

        \State {$switch \leftarrow 0$}
        \While{$Q_{in} \cup Q_{out} \neq \emptyset$} {
            \If{$switch = 0$} {
                \For{counter $\gets 1$ to $|Q_{in}|$}
                {
                    \State  {$(e, s) \leftarrow Q_{in}.pop()$}
                    \lIf{$s \le visit_{in}[e]$} {\State{\textbf{continue}}}
                    \State{$visit_{in}[e] \leftarrow s$}
                    \If{$visit_{out}[e] > result$} {
                        \State{ $result \leftarrow \min(s, visit_{out}[e])$ }
                        \State {\textbf{continue}}
                    }
                    \ForEach{$e' \in N(e) $} {
                        \lIf{$OD(e, e') \le result$}{\State{\textbf{continue}}}
                        \State {$Q_{in} \leftarrow Q_{in} \cup (e', \min(s, OD(e, e')))$}
                    }
                    \State {$switch \leftarrow 1$}
                }
            } \lElse {
                \State {repeat lines 7-18, swap all subscript notations between $in$ and $out$, and change $switch$ back to 0} 
            }
        
        }
        \Return{$result$}

	}
\end{algorithm}

%
Following the above idea, the pseudocode of the online search method is illustrated in Algorithm \ref{alg:BASELINE}.
We start by adding
$(e, \left|e\right|)$ for each $e \in E(u)$ to a priority queue $Q_{out}$ in line 3 according to Corollary~\ref{th:corollary}. Similarly, $Q_{in}$ designed for target $v$ is initialized in the same manner in line 4. The condition in line 6 terminates the search when
all hyperedges reachable from either $u$ or $v$ have been processed. 
The priority-based iteration alternates between $Q_{in}$ and $Q_{out}$ based on the value of $switch$ (0 for $Q_{in}$, otherwise for $Q_{out}$) in line 7. 
In lines 12-14, when exploring a new hyperedge, the algorithm checks if it has been explored from the opposite direction. If so, it indicates the existence of a walk $W(e_u, e_v)$ where $u \in e_u, v \in e_v$, and the result of $MR(u, v)$ should be updated. We prune the search if the current searching branch has no effect on the $MR(u, v)$ result, as demonstrated in lines 10 and 16.

\begin{example}
    Consider the hypergraph shown in Fig. \ref{fig:ex}. When querying $MR(v_1, v_{12})$, We first initialize $Q_{out} = \{(e_1, 2), (e_7,3)\}$ and $Q_{in} = \{(e_3, 3), (e_4, 4)\}$, respectively. When iterating by $Q_{in}$, we first choose $(e_7, 3)$ based on the property of priority queue, and exploring $e_2, e_4$ with $s = 2$. Since the only neighbor of $e_1$ is $e_7$, its traversal will be pruned by line 11. When iterating based on $Q_{out}$ and visit $(e_4, 2)$, since we have $(e_4, 2)$ at the opposite direction, we update the result to 2. We stop the hypergraph traversal because no further exploration can lead to a result greater than 2.   
\end{example}



The worst case for the online MR query searching needs to invoke the whole hypergraph traversal, while the neighbor exploration for a hyperedge takes up to $O(\delta d)$ time complexity.
Formally, we have the following time complexity analysis.

\begin{theorem} 
\label{th:th1} 

The time complexity of Algorithm \ref{alg:BASELINE} is bounded by $O(m  \delta  d)$, where $d = \max_{v \in \mathcal{V}}\left|E(v)\right|$ and $\delta  = \max_{e \in \mathcal{E}}\left|e\right|$.   
\end{theorem}

\textit{Due to the space limitation, the 
proof for Theorem~\ref{th:th1}
and other omitted proofs can be found in Appendix A online~\cite{appendix}.}


\section{Hypergraph labeling Index}
\label{th:sec3.2}

Although the online search method can effectively determine the max-reachability between two vertices, its performance degrades significantly as the graph size increases. 
For example, on the coauth-DBLP dataset with more than 3 million hyperedges, the online algorithm takes more than 60 seconds per query on average (See Section~\ref{sec:6} for details).


To scale for large graphs, in this paper, we aim to design an efficient index-based solution.
In the literature, a dominant index paradigm for reachability problems is the 2-hop labeling framework~\cite{cohen2003reachability}. Its general ideas are outlined as follows: for each vertex $u$, we precompute a set of vertices that can be reached from it and store them in the \textit{label} of $u$.
We refer to the vertices in $u$'s label as the \textit{hubs} of $u$ for presentation simplicity.
The query of reachability between two vertices $u$ and $v$ can be answered based on their labels and will evaluate to true if any of the following conditions hold: $i$) $u$ is in the label of $v$, $ii$) $v$ is in the label of $u$, or $iii$) there exists a common vertex in the labels of $u$ and $v$.

Although the 2-hop labeling framework has undergone comprehensive exploration in recent years \cite{akiba2013fast, abraham2012hierarchical, wang2015efficient,wen2022span}, tailoring this method for our problem is non-trivial.
Intuitively, we can store the \textit{vertex-to-vertex} (VTV) reachability information. To answer $MR(u,v)$, we need to find a set of hub vertices $\mathbb{W}$. For each $w \in \mathbb{W}$, we have $u \stackrel{s_u}\leadsto w$ and $w \stackrel{s_v}\leadsto v$ stored in the index, ``indicating'' $u \stackrel{\min(s_u, s_v)}{\leadsto}  w \stackrel{\min(s_u, s_v)}{\leadsto} v$. 
$MR(u,v)$ can then be obtained by taking the maximum value of such $\min(s_u, s_v)$ as the result.
While intuitive, this approach may lead to false positive results.
One example is shown as follows.


\begin{example} \label{example:vtv}
    Given $\mathcal{H = (V,E)}$ in Figure \ref{fig:ex}(a), vertex $v_1$ reaches vertex $v_3$ via walk $W_1=\{e_7\}$ with $WOD(W_1) = 3$, and vertex $v_{12}$ reaches $v_3$ via walk $W_2=\{e_4\}$ with $WOD(W_2) = 4$. 
    By extending the VTV format 2-hop labeling framework, we have the information $v_1 \stackrel{3}\leadsto v_3$ and $v_{12} \stackrel{4}\leadsto v_{3}$ stored in the index, whereas the information $OD(e_7, e_4)=2$ is missing.
    Based on the index, a query of $MR(v_1, v_{12})$ is at least 3, however, there does not exist a walk $W$ in $\mathcal{H}$ such that $v_1$ reaches $v_3$ via $W$ and $WOD(W)\ge 3$.
\end{example}


\myparagraph{ETE-based index}
One possible way to address this pitfall is to maintain information about the reachability relationship between hyperedges, denoted as \textit{ETE-based index} in this paper.
Specifically, for each hyperedge $e$, we maintain a label set $\mathcal{L}_e(e)$, 
where each label inside $\mathcal{L}_e(e)$ is in the form $(e', s)$, indicating $e \stackrel{s}\leadsto e'$.
The $MR(u, v)$ query can be answered based on the labels of hyperedges $e_u$ and $e_v$ that contain $u$ and $v$, respectively.
Although ETE-based index guarantees correctness, it incurs a high computational cost due to computing the max-reachability between all hyperedge pairs in 
    $e_{pair}(u,v) = \{(e_u, e_v)| u\in e_u \land v\in e_v\}$, as both $u$ and $v$ may appear in multiple hyperedges.

\begin{example}
    Given the hypergraph in Figure~\ref{fig:ex}, and the query $MR(v_4, v_9)$. 
    Under the ETE-based index, this query is answered by taking the maximum max-reachability value between a hyperedge containing $v_4$ and a hyperedge containing $v_9$.
    Given $E(v_4) = \{e_2, e_4, e_7\}$ and $E(v_9) = \{e_3, e_6\}$, we should computing the max-reachability for each of the pairs $(e_2, e_3)$, $(e_2, e_6)$, $(e_4, e_3)$, $(e_4, e_6)$, $(e_7, e_3)$, $(e_7, e_6)$, and return the largest value as the result.
\end{example}

To further speed up this ETE-based query method, we adopt the merge-sort-based algorithm to reduce the time complexity from quadratic to linear, by maintaining and sorting all $e_u \in e_{pair}(u,v)$ in one set and $e_v \in e_{pair}(u,v)$ in another to avoid repeated computation, This approach still needs to incur a significant number of entries.
 
\begin{table}[t]
\centering
\caption{An HL-index $\mathcal{L}$ of $\mathcal{H}$}
\footnotesize
\resizebox{0.98\linewidth}{!}{%
\begin{tabular}{|l|l|l|l|}
\hline
\textbf{$\mathcal{L}(v_1$)} & \textbf{$(e_2, 2), (e_1, 2), (e_7, 3)$} & \textbf{$\mathcal{L}(v_7)$} & \textbf{$(e_2, 6), (e_6, 3)$} \\ \hline
\textbf{$\mathcal{L}(v_2$)} & \textbf{$(e_2, 1), (e_1, 2)$} & \textbf{$\mathcal{L}(v_8)$} & \textbf{$(e_2, 6), (e_6, 3)$} \\ \hline
\textbf{$\mathcal{L}(v_3)$} & \textbf{$(e_2, 6), (e_4, 4), (e_7, 3)$} & \textbf{$\mathcal{L}(v_9)$} & \textbf{$(e_2, 2), (e_3, 3), (e_6, 3)$} \\ \hline
\textbf{$\mathcal{L}(v_4)$} & \textbf{$(e_2, 6), (e_4, 4), (e_7, 3)$} & \textbf{$\mathcal{L}(v_{10})$} & \textbf{$(e_2, 1), (e_5, 3), (e_3, 3)$} \\ \hline
\textbf{$\mathcal{L}(v_5)$} & \textbf{$(e_2, 6), (e_5, 3)$} & \textbf{$\mathcal{L}(v_{11})$} & \textbf{$(e_2, 2), (e_4, 4)$} \\ \hline
\textbf{$\mathcal{L}(v_6)$} & \textbf{$(e_2, 6), (e_5, 3)$} & \textbf{$\mathcal{L}(v_{12})$} & \textbf{$(e_2, 2), (e_4, 4), (e_3, 3)$} \\ \hline
\end{tabular}
}
\label{tab:HLindex}
\end{table}

\myparagraph{VTE-based index}
To alleviate the high computational cost in ETE-based index, 
we then presenting our novel index framework named \underline{H}ypergraph \underline{L}abeling index (HL-index) that maintaining  
\textit{vertex-to-hyperedge} (VTE) reachability relationship information.
Given a hypergraph $\mathcal{H} = (\mathcal{V, E})$
and $u,v\in\mathcal{V}$,
clearly, $u \stackrel{s}{\leadsto} v$ iff there exists a hyperedge $e\in \mathcal{E}$ such that $u \stackrel{s}{\leadsto} e$ and $v \stackrel{s}{\leadsto} e$.
We use a tuple $\{u, e, s\}$ to represent $u \stackrel{s}{\leadsto} e$, and such a tuple is called a \textbf{reachability tuple}. 
HL-index preserves VTE information by maintaining a label set $\mathcal{L}(u)$ for each $u \in \mathcal{V}$. 
Each label in $\mathcal{L}(u)$ is denoted in the form $(e, s)$, indicating vertex $u$ $s$-reaches hyperedge $e$ in $\mathcal{H}$. Compared with VTV, HL-index ensures the correctness of answering $MR$ query. It also significantly reduces the number of entries accessed during the $MR$ query compared with the ETE-based index. A detailed analysis of these three indices is provided in Appendix C \cite{appendix}.
Based on the HL-index, a max-reachability query between $u, v \in \mathcal{V}$ can be correctly answered by the following equation: 
$MR(u, v) = \max\{\min(s_u, s_v)| (e_u, s_u) \in \mathcal{L}(u) \land \ (e_v, s_v) \in \mathcal{L}(v) \land e_u = e_v\}$.

\begin{example}
An HL-index $\mathcal{L}$ of hypergraph in Figure \ref{fig:ex} is shown in Table \ref{tab:HLindex}.
To answer $MR(v_6, v_9)$, we have $\mathcal{L}(v_6)=\{(e_2, 6), (e_5,3)\}$ and $\mathcal{L}(v_9)=\{(e_2, 2),(e_3,3),(e_6, 3)\}$.
There only exists one common hyperedge $e_2$ in both of their labels, thus, $MR(v_6, v_9)=\min(6,2)=2$.


\end{example}

\section{Index Construction}
\label{sec:sec4}
To enable efficient index construction, we first introduce a basic HL-index construction method in Section~\ref{sec:indexcons:base}, which extends the 2-hop labeling framework. We then analyze its limitations.
In Sections \ref{sec:5.2}-\ref{sec:5.3}, we present a fast HL-index construction algorithm integrating two key optimization strategies that address the aforementioned issues.
In Section~\ref{sec:5.4}, we further develop a minimal HL-index construction method.

\subsection{Basic Index Construction Approach}
\label{sec:indexcons:base}

Computing a minimum-size 2-hop index for reachability queries is NP-hard, which has been proven by \cite{cohen2003reachability}. The problem is reducible to the minimum set cover \cite{chvatal1979greedy}, for which a greedy algorithm yields a solution within an $O(\log n)$ factor of optimal, but with prohibitive time costs for large graphs. To improve scalability, \textit{Hierarchy labeling framework}~\cite{akiba2013fast, abraham2012hierarchical} introduces a vertex ordering heuristic that incrementally builds the index via label propagation.
Similarly, computing a minimum HL-index is also NP-hard, since it builds on the 2-hop labeling scheme, we also follow hierarchy labeling by assigning a strict total order over hyperedges, denoted by $\mathcal{O}$, which replaces vertex ordering in the hierarchy, and we adopt the idea of using degree to measure the importance \cite{wen2022span, jin20093}. The hyperedge weight is defined by $\sum_{v \in e}{{\left|E(v)\right|}^2}$ for $e \in \mathcal{E}$. We use the ${\left|E(v)\right|}^2$ to grant higher weight for those hyperedges whose vertices are frequently involved in other hyperedges.
when two hyperedges have the same weight, the one with a smaller ID will precede the other in the hyperedge order. 
Without loss of generality, we say $e_i$ has higher importance compared to hyperedge $e_j$ when $\mathcal{O}(e_i) < \mathcal{O}(e_j)$.

With the hyperedge order, the index construction workflow can be summarized as processing the hyperedges following $\mathcal{O}$ and adding reachability information regarding the current processing hyperedge to the labels of the reachable vertices as needed. 
Before discussing more details, 
we introduce the following concepts that are closely related to an HL-index.



\begin{definition}[Dominant Reachability Tuple] \label{th:def5} 
    Given a vertex $u$ and a hyperedge $e$, a reachability tuple $\{u, e, s\}$ dominates a tuple $\{u, e, s'\}$ if $s > s'$. $\{u, e, s\}$ is a dominant reachability tuple if it is not dominated by any other tuples.
\end{definition}




\begin{definition}[Transitive Cover]\label{th:def6}

    Given a reachability tuple $\{u, e, s\}$, it is transitively covered by a hyperedge $e_w$ if $i)$ $u \stackrel{s'}{\leadsto} e_w$, $e_w \stackrel{s''}{\leadsto} e$, $s'\ge s$, $s'' \ge s$, and $ii)$ $\mathcal{O}(e_w) < \mathcal{O}(e)$.
    In this case, $e_w$ is called an intermediate hyperedge of $\{u, e, s\}$.
\end{definition}



\begin{example}
    As shown in Figure \ref{fig:ex}(a), $v_1 \stackrel{2}{\leadsto} e_4$ through $W_1 = \{e_7, e_4\}$ and $v_1 \stackrel{1}{\leadsto} e_4$ through $W_2 = \{e_7, e_2, e_5, e_3, e_4\}$. In this case, we have the reachability tuple $\{v_1, e_4, 2\}$ dominates the reachability tuple $\{v_1, e_4, 1\}$. 
    Suppose we have $\mathcal{O}(e_2) < \mathcal{O}(e_6) < \mathcal{O}(e_4)$, the reachability tuple $\{v_9, e_4, 2\}$ is transitively covered by hyperedge $e_2$, since $v_9 \stackrel{2}{\leadsto} e_2$ through $W_3 = \{e_6, e_2\}$ and $e_2 \stackrel{2}{\leadsto} e_4$ through $W_4 = \{e_2, e_4\}$. 
\end{example}

Now, we can formalize when reachability information (represented by reachability tuples) is considered not needed. 
Such information should be ignored and excluded from the index.
Firstly, it is clear that information carried by non-dominant reachability tuples can be safely omitted.
Secondly, consider the case in a hypergraph $\mathcal{H}$ when a hyperedge $e_w$ is of the highest importance among all hyperedges that transitively cover a dominant reachability tuple $\{u, e, s\}$. That is, $u \stackrel{s_1}{\leadsto} e_w$, $e_w \stackrel{s_2}{\leadsto} e$, $s=\min(s_1,s_2)$.
Suppose the max-reachability $u \stackrel{k}{\leadsto} v$ can be derived from $u \stackrel{s}{\leadsto} e$ and $e \stackrel{s_3}{\leadsto} v$, then it can also be derived from $u \stackrel{s_1}{\leadsto} e_w$ and $e_w \stackrel{\min(s_2,s_3)}{\leadsto} v$. 
Note that, there does not exist a hyperedge that transitively covers $\{v, e_w, s_3\}$; otherwise, $e_w$ is not of the highest importance.
Therefore, $\{u, e, s\}$ contains redundant reachability information and can be omitted.
On this basis, we formally define the following core concept in computing the HL-index.

\begin{definition} [Essential Tuple]\label{def:Essential}
    A reachability tuple $\{u, e, s\}$ is an essential tuple if $i)$ $\{u, e, s\}$ is dominant, and $ii)$ there does not exist a hyperedge $e_w$ that transitively covers $\{u, e, s\}$.
\end{definition}

Following the definition of the essential tuple, the task of finding an HL-index can be transferred to efficiently compute a set of essential reachability tuples that preserve the max-reachability information between all pairs of vertices in a hypergraph.
We process hyperedges in descending importance order; for each hyperedge 
$e$, the algorithm traverses walks starting from 
$e$, extracts reachability tuples, and adds labels when they are verified as essential.
We then introduce the methods of computing dominant tuples, verifying essential tuples, and propose a basic index construction algorithm.
\begin{lemma}\label{th:visitv}



When processing a hyperedge $e \in \mathcal{E}$, 
given a set of essential tuples $\mathcal{Q}^*$ and a set of discovered reachability tuples $\mathcal{Q}$ that have not been verified as essential, a reachability tuple $\{u,e,s\} \in \mathcal{Q}$ is dominant if $i)$ $\{u,e,s\}$ is not dominated by any essential tuples in $\mathcal{Q}^*$, and $ii)$ $s$ is the largest among all reachability tuples in $\mathcal{Q}$.
\end{lemma}

Based on Lemma~\ref{th:visitv}, we apply a variant of Dijkstra’s algorithm, akin to the online search in Algorithm~\ref{alg:BASELINE}, to discover dominant reachability tuples for each $e \in \mathcal{E}$. A priority queue maintains frontier items $(e_u,s)$, denoting $e \stackrel{s}{\leadsto} e_u$ and $u \stackrel{s}{\leadsto} e$ for all $u \in e_u$, with the highest-$s$ item popped first. A tuple $\{u,e,s\}$ is discovered when $(e_u,s)$ enters the frontier, and $u$ is considered visited by $e$ when $(e_u,s)$ is removed. Dominance holds at the first visit of $u$, while later traversals to the same $e_u$ can be pruned as they yield no new dominant tuples.

Next, before discussing the verification of essential tuples, we further investigate the transitive-covering relationships in hypergraphs.
In Definition~\ref{th:def6}, we define the transitive-covering relationship between vertex and hyperedge. As an extension, we also define the transitive-covering relationship between hyperedges as follows. 
Given hyperedges $e_u$, $e_v$ and $e_w$, we say $e_u \stackrel{s}{\leadsto} e_v$ is transitively covered by an intermediate hyperedge $e_w$ if $i)$ $e_u \stackrel{s'}{\leadsto} e_w$, $e_w \stackrel{s''}{\leadsto} e_v$, $s',s'' \ge s$, and $ii)$ $\mathcal{O}(e_w) < \mathcal{O}(e_u) , \mathcal{O}(e_v)$. 
We observe a connection between these two transitive-covering relationships, as detailed below.


\begin{lemma}\label{lem:transitive-covering}
Given $\{u, e, s\}$ derived by a hyperedge $e_u \in E(u)$ s.t. $e \stackrel{s}{\leadsto} e_u$ and $\mathcal{O}(e)<\mathcal{O}(e_u)$, a hyperedge $e_w$ transitively covers such $\{u, e, s\}$ iff $e_w$ transitively covers $e \stackrel{s}{\leadsto} e_u$.
\end{lemma}

Based on Lemma~\ref{lem:transitive-covering}, the task of verifying whether a dominant tuple $\{u, e, s\}$ is essential, i.e., checking the existence of an intermediate hyperedge $e_w$ of $\{u, e, s\}$, can be transferred to determining whether there exists an intermediate hyperedge $e_w$ that transitively covers $e \stackrel{s}{\leadsto} e_u$ for any $e_u$ such that $u \in e_u$ and $\mathcal{O}(e)<\mathcal{O}(e_u)$.
Moreover, we can further reduce
the number of transitive covering checks during the search for reachability tuples by using the following pruning rule.

\begin{lemma}
\label{th:lemma3} 
Given a reachability tuple $\{u, e, s\}$, it is transitively covered if there exists a walk $W(e, e_u)$ such that $u \in e_u$, $WOD(W(e, e_u)) \geq s$ and $W(e, e_u)$ contains at least one hyperedge with higher importance compared to $e$.
\end{lemma}




\begin{algorithm}[t]
	{
		\SetVline
		\footnotesize
            \caption{{Basic HL-index Construction Method}}
		\label{alg:alg3}
		\Input{A hypergraph $\mathcal{H=(V, E)}$}
		\Output{An HL-index $\mathcal{L}$}
     \State{$\mathcal{L}(u) \gets \emptyset$, $visited[u] \gets$ null \textbf{for each} $u\in \mathcal{V}$ }
    \State{$visited_e[e] \gets$ null \textbf{for each} $e\in \mathcal{E}$ }

            \ForEach{$e \in \mathcal{E}$ \textup{in descending order of its importance}} {
                
                \State{$Q \leftarrow$ a priority queue with $(e, \left|e\right|)$}
                \While{$Q \neq \emptyset$} {
                    \State {$(e_u, s) \leftarrow Q.pop()$}
                    \State{$visited_e[e_u] \gets e$}
                    \lIf{$\exists{e_w} : \mathcal{O}(e_w) < \mathcal{O}(e) \land e_w \stackrel{s}\leadsto e \land e_w \stackrel{s}\leadsto e_u$} {\State{\textbf{continue}}}
        
                    \ForEach{$u \in e_u$} {
                        
                        \lIf{$visited[u] = e$} {\State{\textbf{continue}}}
                    \State {$\mathcal{L}(u) \leftarrow \mathcal{L}(u) \cup \{(e, s)\}$}
                    \State {$visited[u] \leftarrow e$}
                    }
    
                    \ForEach{$e_v \in N(e_u)$} {
                        \lIf{$\mathcal{O}(e_v) \le \mathcal{O}(e)$}{\State{\textbf{continue}}}
                        \lIf{$visited_e[e_v] = e$}{\State{\textbf{continue}}}
                        \State {$Q \leftarrow Q \cup \{(e_v, \min(s, OD(e_u, e_v)))\}$}
                    }
                }
            }

            \Return{$\mathcal{L}$}
	}

\end{algorithm}

With Lemmas~\ref{th:visitv}–\ref{th:lemma3} for dominant tuple computation, transitive covering, and essential tuple verification, we design the basic index construction in Algorithm~\ref{alg:alg3}. We first pre-compute a hyperedge order by importance in $O(m\cdot(d\delta+\log m))$ time complexity. For each hyperedge $e$ (lines 4–16), we compute its essential tuples. Based on Lemma~\ref{th:visitv}, we track visits of hyperedges and vertices (lines 7 and 12) to detect dominant tuples and prune redundant searches (lines 10 and 15). At line 8, we check whether $e \stackrel{s}{\leadsto} e_u$ is transitively covered. If so, further traversal can be pruned since it will not contribute to essential tuples. Since the VTE-based HL-index lacks sufficient information for this detection, we implement it by online search: if existing index entries imply the $s$-reachability between $u \in e_u$ and $v \in e$, we perform a bidirectional BFS from $e$ and $e_u$ to detect if any other hyperedge covers $e\stackrel{s}{\leadsto}e_u$. If not, by Definition~\ref{def:Essential} and Lemma~\ref{lem:transitive-covering}, those $\{u,e,s\}$ for $u \in e_u$ with unvisited $u$ are essential and added to $\mathcal{L}(u)$ (line 11). Lines 13–16 then continue exploration, ignoring neighbors of importance no smaller than $e$, as such walks cannot yield new essential tuples according to Lemma \ref{th:lemma3}.



\begin{theorem} [Correctness Analysis] \label{th:correctness}
Given the HL-index $\mathcal{L}$ constructed by Algorithm~\ref{alg:alg3} for a hypergraph $\mathcal{H} = (\mathcal{V, E})$, the max-reachability queries between any pair of vertices $u,v \in \mathcal{V}$ can be correctly determined through  $\mathcal{L}(u)$ and $\mathcal{L}(v)$. 
\end{theorem}

\myparagraph{Limitations}
Although Algorithm \ref{alg:alg3} effectively builds the HL-index, it is not efficient due to the following three limitations.


\begin{itemize}[leftmargin=*]

\item {\underline{Limitation 1: Inefficient transitive covering detection.}} 
One challenge is to efficiently determine whether the reachability relationship between hyperedges $e$ and $e_u$ has been transitively covered by another hyperedge $e_w$ (line 8 of Algorithm \ref{alg:alg3}).
As mentioned before, our VTE-based HL-index does not provide adequate information to effectively detect hyperedge transitive covering, and line 8 is implemented as a bidirectional BFS from $e$ and $e_u$, respectively, to find a hyperedge $e_w$ that can transitively cover $e \stackrel{s}{\leadsto} e_u$.
This online search approach is effective, yet far from efficient, as it necessitates traversing the hypergraph frequently.

\item {\underline{Limitation 2: Redundant hyperedge neighbor computation.}} 
In lines 13-16 of Algorithm \ref{alg:alg3}, we need to compute $N(e_u)$ and the corresponding $OD(e_u, e_v)$ for every $e_v \in N(e_u)$ for further traversal.
For an arbitrary hyperedge $e_u$, suppose $k$ is the number of hyperedges $e \in \mathcal{E}$ such that the construction procedure from $e$ involves a traversal through $e_u$. 
Then, if $N(e_u)$ is computed on the fly, the same information will be computed up to $k$ times.
Let $k_{avg}$ denote the average value of $k$ across all hyperedges, the total time complexity related to real-time neighbor detection is bounded by $O(m  k_{avg}  \delta  d)$, 
indicating high computational costs.
Furthermore, maintaining $N(e)$ for all $e \in \mathcal{E}$ in an adjacent list is memory expensive, since its space complexity is up to $O(m  \delta  d)$. 

\item {\underline{Limitation 3: Non-minimal index size.}} 
A good HL-index $\mathcal{L}$ should comply with the notion of minimality,  including $i)$ (Completeness) $MR$ queries between any pair of vertices could be correctly answered using only labels in $\mathcal{L}$, and $ii)$ (Necessity) if any label is removed from $\mathcal{L}$, we will fail to correctly answer at least one $MR$ query.
Although the HL-index from Algorithm~\ref{alg:alg3} satisfies the completeness according to Theorem~\ref{th:correctness}, it does not guarantee the necessity. 
The reason is analyzed as follows: In our VTE-based HL-index, it is possible to remove some hyperedges without compromising the correctness of $MR$ queries. 
Unlike in the construction of a conventional VTV-based index, it is hard to verify if a reachability tuple adheres to the necessity when it is discovered.
For a VTV-based index, when the information of reachability from hub vertex $w$ to a vertex $u$ (denoted as $w \leadsto u$) is discovered,
a query for reachability between $u$ and $w$ can be performed using the current index, 
and the query result not only determines whether this information $w \leadsto u$ is necessary for $(u,w)$ query, but also determines its necessity for $(u,v)$ queries for all $v$.
However, for VTE-based index, 
given two reachability tuples $\{u,e,s_u\}$, $\{v,e,s_v\}$ where $s_v < s_u$, the necessity of $\{u,e,s_u\}$ for $MR(u,v)$ query highly depends on $\{v,e,s_v\}$ while the later has not been discovered when we discover $\{u,e,s_u\}$, due to the usage of the priority queue.

\end{itemize}
To address limitations 1 and 2, we first explore the alternative method for checking hyperedges covering relationship in Section-\ref{sec:5.2}. We then introduce an auxiliary data structure in Section-\ref{sec:5.3} that dynamically maintains hyperedge neighbor information that is in active use.
To rectify limitation 3, we develop a novel strategy to efficiently determine all necessary labels for a minimal HL-index, as discussed in Section-\ref{sec:5.4}.


\subsection{Accelerating Transitive Cover Detection} \label{sec:5.2}


A naive solution to detect hyperedge transitive covering is to build an additional ETE-based index for tracking hyperedge reachability. However, this incurs substantial redundancy. We propose a novel alternative that leverages properties of undirected graphs to eliminate explicit covering checks. We begin by introducing the notion of maximum cover degree.

    \begin{definition}[Maximum Cover Degree]
    \label{th:def7}
   Given a hyperedge $e$, its maximum cover degree refers to the WOD degree among all walks starting at $e_w$ and ending at $e$, where $\mathcal{O}(e_w) < \mathcal{O}(e)$, denoted as $MCD(e) = \max_{e_w \in \mathcal{E}}\{WOD(W(e_w, e)) | \mathcal{O}(e_w) < \mathcal{O}(e) \land W(e_w, e) \in \mathcal{W}(e_w, e)\}$. 
    
\end{definition}

\begin{lemma}\label{th:lemma6} 
Given two hyperedges $e$ and $e_u$, where $e\stackrel{s}{\leadsto} e_u$ and $\mathcal{O}(e) < \mathcal{O}(e_u)$, 
there exists a hyperedge $e_w$ with $\mathcal{O}(e_w) < \mathcal{O}(e)$ that transitively covers $e\stackrel{s}{\leadsto} e_u$ iff $MCD(e) \ge s$.
\end{lemma}

According to Lemma \ref{th:lemma6}, during the HL-index construction procedure from a hyperedge $e$, the task of verifying whether there exists a hyperedge $e_w$ with higher importance covers $e \stackrel{s}{\leadsto} e_u$, can be simplified as comparing $s$ with $MCD(e)$. 
Although naively enumerating all walk $W(*, e)$ to compute $MCD(e)$ is costly, we notice that $MCD(e)$ can be obtained without exploring all walks ending at $e$, as discussed below.

\begin{lemma} \label{lemma:MCD}
    The $MCD$ value of every hyperedge $e$ can be correctly computed by the walks enumerated in Algorithm \ref{alg:alg3}.
\end{lemma}

Based on Lemma \ref{lemma:MCD},  walks that contribute to $MCD(e)$ will always be explored during the index construction process. 
Therefore, the lower bound of $MCD(e)$ can be conveniently updated by monitoring the highest walk overlapping degree of all the walks ended at $e$ during the index construction procedure from those hyperedges with higher importance than $e$. Since the index construction follows the descending order of hyperedge importance, $MCD(e)$ is the same as its lower bound when start index construction from $e$. Therefore, hyperedge transitive cover detection can be performed both efficiently and effectively during HL-index construction.

\begin{example}
    Consider a sub-hypergraph $\mathcal{H'} = (\mathcal{V}, \mathcal{E'})$ of the hypergraph $\mathcal{H}$ in Figure \ref{fig:ex}(a), where $\mathcal{E'} = \{e_1, e_2, e_4, e_7\}$ and $\mathcal{O}(e_2) < \mathcal{O}(e_4) < \mathcal{O}(e_7) < \mathcal{O}(e_1)$.
    During the construction procedure from $e_2$, we update lower bound of $MCD(e_4)$ to 2 after we obtain a walk $\{e_2, e_4\}$. 
    Similarly, the lower bound $MCD$ values of $e_7$, $e_1$ are set to 2 and 1, respectively.
    At the beginning construction stage of $e_4$, we assign $MCD(e_4)$'s lower bound to $MCD(e_4)$, which is 2, and it will prune the walk moving on to $e_7$ with $W = \{e_4, e_7\}$ and $WOD(W) = 2$, which is no larger than $MCD(e_4)$. This is because the information of $W$ is covered by another walk $W(e_4, e_2) \oplus W(e_2, e_4) \oplus W$, where both $WOD(W(e_4, e_2)), WOD(W(e_2, e_4)) = MCD(e_4) \ge 2$.

    
\end{example}


\subsection{Reducing neighbor Computation} \label{sec:5.3}


As maintaining $N(e)$ for all $e \in \mathcal{E}$
suffers heavy memory consumption, to avoid repeatedly computing neighbors for each hyperedge during the index construction procedure, we propose a novel lightweight neighbor-index, denoted as $\mathcal{M}$. 
For a hyperedge $e$, each element in $\mathcal{M}(e)$ is a neighbor tuple $(e', s)$, where $e' \in N(e)$ and $s = OD(e,e')$.
$\mathcal{M}(e)$ will be initialized at the first time $e$ is traversed and will only maintain all necessary neighbor information for later index construction stages. In other words, for each $e$, $N(e)$ will be computed exactly once, stored in $\mathcal{M}(e)$. Every $(e', s) \in \mathcal{M}(e)$ with no future access will be dynamically removed to reduce the memory cost, supported by the following lemma.


\begin{lemma} \label{th:lemma10}
    Given hyperedges $e$, $e'$, $e_u$ where $\mathcal{O}(e)<\mathcal{O}(e')<\mathcal{O}(e_u)$, $e_v \in N(e_u)$ and $\mathcal{O}(e')<\mathcal{O}(e_v)$. If $e \stackrel{k}{\leadsto} e_u$ and $OD(e_u,e_v)\le k$, then the relationship $e' \stackrel{s}{\leadsto} e_v$ derived from $W(e',e_v)=W(e',e_u) \oplus \{e_v\}$, is transitively covered by $e$.
\end{lemma}

According to Lemma \ref{th:lemma10}, suppose we are performing two consecutive index construction procedures from hyperedges $e$ and $e'$. 
If we have traversed to a hyperedge $e_u$ from $e$ with $WOD(W(e, e_u))=k$ and explored a hyperedge $e_v \in N(e_u)$ with $OD(e_u,e_v)\le k$, a traversal from $e'$ to $e_u$ and then visits $e_v$ can be safely terminated. This is because the reachability information carried by such walk $W(e',e_v)$ is covered by $e$.
Similarly, if we traverse to $e_v$ from $e$ with $WOD(W(e, e_v))=OD(e_u,e_v)$, future traversal from $e'$ to $e_u$ through $e_v$ can also be safely terminated.
Therefore, the neighbor information between $e_u$ and $e_v$ is redundant when constructing from hyperedges that have no higher importance than $e$ and hence can be removed from $\mathcal{M}(e_u)$ and $\mathcal{M}(e_v)$.
To support efficient neighbor tuple deletion, for each hyperedge $e$, neighbor tuples in $\mathcal{M}(e)$ are arranged in descending order based on the importance of their first element.

\begin{algorithm}[t]
{
    \SetVline
    \footnotesize
    \caption{Fast HL-index Construction Method} 
    
    \label{alg:alg4}
    \Input{A hypergraph $\mathcal{H=(V, E)}$}
    \Output{An HL-index $\mathcal{L}$}
    \State{$\mathcal{L}(u) \gets \emptyset$, $visited[u] \gets$ null \textbf{for each} $u\in \mathcal{V}$ }
    \State{$MCD(e) \gets 0$, $\mathcal{M}(e)\gets$ null, $visited_e[e] \gets$ null \textbf{for each} $e \in \mathcal{E}$}


    \ForEach{$e \in \mathcal{E}$~\textup{in descending order of its importance}}
    {
        \lIf{$MCD(e) = \left|e\right|$}{\State{\textbf{continue}}}
        \State {$Q \leftarrow$ a priority queue with $(e, \left|e\right|)$}
        \While{$Q \neq \emptyset$} {
            \State{$(e_u, s) \gets Q.pop()$}
            \State{$visited_e[e_u] \gets e$}
            \State {$MCD(e_u) \leftarrow \max(s, MCD(e_u))$}
          
            \ForEach{$u \in e_u$} 
            { 
                \lIf{$visited[u] = e$}{\State{\textbf{continue}}}
                \State {$\mathcal{L}(u) \leftarrow \mathcal{L}(u) \cup \{(e, s)\}$}
                \State {$visited[u] \leftarrow e$}
            }
            \If{$\mathcal{M}(e_u)$ \textup{has not been initialized yet}} {
                \StateCmt{$\mathcal{M}(e_u) \gets \emptyset$}{compute $N(e_u)$}
                \For{$e_v \in N(e_u)$} {
                    \lIf{$\mathcal{O}(e_v) \le \mathcal{O}(e)$}{\State{\textbf{continue}}}
                    \State{$\mathcal{M}(e_u) \leftarrow \mathcal{M}(e_u) \cup \{(e_v, OD(e_u, e_v))\}$}
                }
            }
    
            \ForEach{$(e_v, s') \in \mathcal{M}(e_u)$} {
                \If {$s' > MCD(e) \land visited_e[e_v] \neq e$}{\State{$Q \leftarrow Q \cup \{(e_v, \min(s,s'))\}$}}
                \If{$s' \le s$} {
                    \State{$\mathcal{M}(e_u) \leftarrow \mathcal{M}(e_u) \setminus \{(e_v, s')\}$}
                    \State{$\mathcal{M}(e_v) \leftarrow \mathcal{M}(e_v) \setminus \{(e_u, s')\}$}
                }
            }
             
        }
    }      
    \Return{$\mathcal{L}$}
}

\end{algorithm}

\myparagraph{Fast HL-index construction algorithm} Incorporating the techniques proposed in Sections~\ref{sec:5.2}-\ref{sec:5.3}, we illustrate the details of the fast HL-index construction algorithm in Algorithm \ref{alg:alg4}. 
It accelerates the index construction and follows a similar workflow as Algorithm \ref{alg:alg3} with two key differences.
\underline{First}, $MCD$ is utilized in transitive covering detection.
Line 9 updates the lower bound of the $MCD$ value of a hyperedge $e_u$ each time a walk visits $e_u$, while the $MCD$ value of hyperedge $e$ is the same as its lower bound when we start constructing from $e$. 
Moreover, the transitive covering detection is performed prior to $(e_u,s)$ being pushed onto $Q$, which involves comparing the overlapping degree of the current walk with $MCD(e)$ (line 20) based on Lemma~\ref{th:lemma6}.
\underline{Second}, we maintain the neighbor-index $\mathcal{M}$ to reduce neighbor computation, which consists of two phases:
$i)$ The initialization of $\mathcal{M}(e_u)$ (lines 14-18) occurs when $e_u$ is visited for the first time throughout the whole index construction procedure.
We compute the corresponding $N(e_u)$ and insert information for each neighbor $e_v$ of $e_u$ into $\mathcal{M}(e_u)$ with an exception when $e_v$ has importance no smaller than the current processing $e$. This exclusion is supported by Lemma ~\ref{th:lemma3}, as shown in line 17.
$ii)$ The update phase is shown in lines 22-24, which removes redundant neighbors that will not be further used (Lemma \ref{th:lemma10}). Lines 20-22 demonstrated the procedures for traversals. Since we ensure the current $s$ is greater than  $MCD(e)$, we only need to compare  $s'
$ with $MCD(e)$ to perform the transitive cover detection in line 20
, similar to line 8 of Algorithm \ref{alg:alg3}.

Let $l = \sum_{u\in \mathcal{V}}(\left|\mathcal{L}(u)\right|)$. Every tuple $\{u, e, s\}$ where $(e, s) \in \mathcal{L}(u)$ may lead to the $d$ times push operation to the queue (i.e., all $e_u \in E(u)$ are visited during the construction of $e$, so there are up to $l \cdot d$ items inside the queue.
 Furthermore, suppose $\alpha_{e}$ is the maximum peak size of $\mathcal{M}(e)$ among all $e \in \mathcal{E}$ during HL-index construction, i.e., $\alpha_{e} = \max_{e\in \mathcal{E}}(\left|\mathcal{M}(e)\right|)$, and we use $\alpha$ to denote the total number of elements inserted into $\mathcal{M}$.  For a hyperedge $e$ and the corresponding neighbor-index $\mathcal{M}(e)$, only those $e_v$ with $OD(e, e_v) > MCD(e)$ will be maintained in $\mathcal{M}(e)$, so we have $\alpha_{e} << \eta_{max}$. 
 Based on these, the complexity results can be formally stated below.

\begin{theorem} [Time Complexity] 
\label{the:advan:timecom}
    The time complexity of Algorithm~\ref{alg:alg4} is $O(l  d \cdot (\log(l d) + \delta + \alpha_{e}) + m \delta d   + \alpha \log\alpha_{e})$.
\end{theorem}

\begin{theorem} [Space Complexity] \label{theo:A3space}
    The HL-index size computed by either Algorithm~\ref{alg:alg4} or Algorithm~\ref{alg:alg3}, is bounded by $O(\sum_{u \in \mathcal{V}}(|\mathcal{E}_{\leq u}|))$ where
    $\mathcal{E}_{\leq u} = \{e| e_u \in E(u) \land \mathcal{O}(e) \leq \mathcal{O}(e_u)\}$.
    The space complexity of neighbor-index $\mathcal{M}$ in Algorithm~\ref{alg:alg4} is bounded by $O(m  \alpha_{e})$, where $\alpha_{e} << \eta_{max}$.

\end{theorem}

\subsection{Generating Minimal HL-index} \label{sec:5.4}

As mentioned in Section~\ref{sec:indexcons:base}, the HL-index $\mathcal{L}$ constructed by Algorithm \ref{alg:alg4} does not necessarily satisfy the necessity property in minimal requirement.
Thus, we propose a novel method for minimal HL-index generation in this subsection.
Given a complete HL-index $\mathcal{L}$ of a hypergraph $\mathcal{H}=(\mathcal{V}, \mathcal{E})$, a hyperedge $e \in \mathcal{E}$ and two vertices $u,v \in \mathcal{V}$. If there exist two labels $(e, s_u) \in \mathcal{L}(u)$, $(e, s_v) \in \mathcal{L}(v)$, we say $u,v$ are reachable vertices through $e$ and $e$ supports $u {\leadsto} e {\leadsto} v$.
Consider a dual structure $\mathcal{D}$ of the complete HL-index $\mathcal{L}$, that is, for each $e \in \mathcal{E}$, $\mathcal{D}(e)=\{(u,s_u)| (e, s_u) \in \mathcal{L}(u)\}$. 
Evidently, the set of vertices reachable through $e$ is $\mathcal{V}(\mathcal{D}(e))=\{u| (e, s_u) \in \mathcal{L}(u)\}$.
For a hyperedge $e' \neq e$, if there exist two labels $(e', s'_u) \in \mathcal{L}(u)$, $(e', s'_v) \in \mathcal{L}(v)$ such that $\min(s'_u, s'_v) \geq \min(s_u, s_v)$, we also consider $e'$ a supporting hyperedge of reachability $u {\leadsto} e {\leadsto} v$.
Clearly, a label $(e, s) \in \mathcal{L}(u)$ can be safely removed from $\mathcal{L}$ without impacting the completeness of $\mathcal{L}$ if 
for each vertex $v \in \mathcal{V}(\mathcal{D}(e))$, there exists a hyperedge $e' \neq e$ that supports $u {\leadsto} e {\leadsto} v$. We say the label and its corresponding essential tuple $\{u, e, s\}$ are redundant in this case.
Intuitively, to obtain a minimal HL-index $\mathcal{L}^*$, we can iteratively identify and remove redundant labels one at a time from the current $\mathcal{L}$ until no redundant label exists. 
We then introduce several optimizations to speed up the processing.

\begin{lemma}\label{lem:pairredundant}
    If $MR(u, v)$ is only supported by $e$, then two entries $(e, s_u) ,(e, s_v) \in \mathcal{L}(u),\mathcal{L}(v)$ are not redundant.
\end{lemma}

Based on Lemma~\ref{lem:pairredundant}, 
a beneficial strategy is to group essential tuples by their hyperedges and verify essential tuples of the same hyperedge sequentially since they are closely related concerning redundancy.
During the verification of an essential tuple $\{u, e, s_u\}$, i.e. when processing $u$, once the reachability $u {\leadsto} e {\leadsto} v$ is found not supported by any hyperedges other than $e$, we can conclude essential tuples $\{u, e, s_u\}$ and $\{v, e, s_v\}$ are not redundant.
Later, when verifying another essential tuple $\{w, e, s_w\}$, since the supporting check for reachability $w {\leadsto} e {\leadsto} u$ with a previously processed $u$ has already been performed, we only need to check whether the reachability $w {\leadsto} e {\leadsto} v$ is supported by hyperedges other than $e$ for those unprocessed vertices $v \in \mathcal{V}(\mathcal{D}(e))$.
Alternatively, the verification of an essential tuple $\{w, e, s_w\}$ can be implemented by maintaining
a vertex set $\mathcal{S} \subseteq \mathcal{V}(\mathcal{D}(e))$ such that, for each vertex $v \in \mathcal{S}$, the reachability $w {\leadsto} e {\leadsto} v$ is supported by a hyperedge $e' \in \{e''| (e'',s) \in \mathcal{L}(w) \land e'' \neq e\}$. Then, $\{w, e, s\}$ can be determined not redundant if any of the two conditions holds: $i)$ $\mathcal{S}$ is missing some unprocessed vertices in $\mathcal{V}(\mathcal{D}(e))$, or $ii)$ $w {\leadsto} e {\leadsto} u$ with a previously processed $u$ is not supported by any hyperedge other than $e$.
This verification process can be further improved based on the following observation.




\begin{observation}\label{ob:verifyorder}
Given two essential tuples $\{u, e, s_u\}$, $\{v, e, s_v\}$ where $s_u \geq s_v$, $u {\leadsto} e {\leadsto} v$ is supported by $e'\neq e$, if and only if 
$(e', s'_u) \in \mathcal{L}(u)$, $(v, s'_v) \in \mathcal{D}(e')$ and $s'_u,s'_v \geq s_v$.
\end{observation}

Observation~\ref{ob:verifyorder} suggests that if we enforce a non-ascending order of $s$ on the essential tuples $\{*, e, s\}$ of $e$ and verify them from the highest $s$ to the lowest, during the verification of an essential tuple $\{u, e, s_u\}$, all unverified essential tuples $\{v, e, s_v\}$ will have $s_v \leq s_u$, and we can efficiently avoid unnecessary support checking by filtering out $(e', s'_u) \in \mathcal{L}(u)$ with $s'_u < s_v$ and $(v, s'_v) \in \mathcal{D}(e')$ with $s’_v < s_v$.
These optimizations help identifying redundant labels, forming the basis for the algorithm to generate a minimal HL-index.

\begin{algorithm}[t]
{
    \SetVline
    \footnotesize
    \caption{Minimal HL-index Generation} 
    
    \label{alg:minimalnew}
    \Input{A complete HL-index $\mathcal{L}$ of a hypergraph $\mathcal{H}=(\mathcal{V}, \mathcal{E})$, a dual data structure $\mathcal{D}$ of $\mathcal{L}$}
    \Output{A minimal HL-index $\mathcal{L}^*$ of $\mathcal{H}$}
    \State{$\mathcal{L}^*(u) \gets \emptyset$ \textbf{for each} $u\in \mathcal{V}$}

    \For{$e \in \mathcal{E}$~\textup{in descending order of its importance}} {
        \State{$\mathcal{I}(e)\gets \emptyset$ \textbf{for each} $e \in \mathcal{E}, \mathcal{NR} \gets \emptyset$}

        \For{$(v, s_v) \in \mathcal{D}(e)$} {
            \For{$(e', s'_v) \in \mathcal{L}(v)$} {
                \lIf{$s'_v \ge s_v$} {
                    \State{$\mathcal{I}(e') \leftarrow \mathcal{I}(e') \cup \{(v, s_v)\}$}
                    
                }
            
            }

        }
        \For{$(u, s_u) \in \mathcal{D}(e)$~\textup{in non-ascending order of $s_u$}} {
            \State{$\mathcal{S} \gets \emptyset$}

            \For{$(e', s'_u) \in \mathcal{L}(u)$ s.t. $e' \neq e$} {
                \For{$(v, s_v) \in \mathcal{I}(e')$} {
                    \lIf{$v \notin \mathcal{V}(\mathcal{D}(e)) \lor s'_u < s_v$}{\State{\textbf{continue}}}
                    \State{$\mathcal{S} \leftarrow \mathcal{S} \cup \{v\}$}
                    \lIf{$\left|\mathcal{S}\right| = \left|\mathcal{D}(e)\right|$}{\State{\textbf{goto} line 14}}
                }
                
            }
            \If{$\left|\mathcal{S}\right| < \left|\mathcal{D}(e)\right| \lor u \in \mathcal{NR}$ } {
                \State{$\mathcal{L}^*(u) \gets \mathcal{L}^*(u) \cup \{(e, s_u)\}$}
                \State{$\mathcal{NR} \leftarrow \mathcal{NR} \cup \{w\}$ \textbf{for each} $w \in \mathcal{V}(\mathcal{D}(e)) \setminus \mathcal{S}$}
                
            } \Else {
                \State{$\mathcal{L}(u) \leftarrow \mathcal{L}(u) \setminus \{(e, s_u)\}$}
            }
            \State{$\mathcal{NR} \leftarrow \mathcal{NR} \setminus \{u\}$, $\mathcal{D}(e) \leftarrow \mathcal{D}(e) \setminus \{(u, s_u)\}$}
            \lIf{$\left|\mathcal{NR}\right| = \left|\mathcal{D}(e)\right|$} {\State{\textbf{break}}}
        }
        \State{$\mathcal{L}^*(u) \leftarrow \mathcal{L}^*(u) \cup \{(e, s_u)\}$ \textbf{for each} $(u, s_u) \in \mathcal{D}(e)$}
    }
    \Return{$\mathcal{L}^*$}
}

\end{algorithm}

\myparagraph{Minimal HL-index generation} 
Algorithm \ref{alg:minimalnew} presents the pseudocode for generating a minimal HL-index $\mathcal{L}^*$ from the HL-index $\mathcal{L}$ produced by Algorithm \ref{alg:alg4}. 
The corresponding dual data structure $\mathcal{D}$ can be computed along with $\mathcal{L}$ without extra cost.
For each hyperedge $e$, before we verify its essential tuples, we initialize two auxiliary data structures, the inverted set $\mathcal{I}$ and the non-redundant set $\mathcal{NR}$ in lines 3-6.
Based on Observation~\ref{ob:verifyorder}, 
for each potential supporting hyperedge $e'$, we only need to consider the support checking for those $v$ from a subset of $\mathcal{V}(\mathcal{D}(e')) \cap \mathcal{V}(\mathcal{D}(e))$. $v$ should satisfy that $(v, s_v) \in \mathcal{D}(e)$, $(e', s'_v) \in \mathcal{L}(v)$, $s'_v \ge s_v$, and we store such $v$ together with its overlapping degree $s_v$ regarding $e$ in $\mathcal{I}(e')$. According to Lemma~\ref{lem:pairredundant}, when $\{u, e, s_u\}$ is being verified as non-redundant, some of the unverified essential tuples $\{w, e, s_w\}$ can be determined not redundant simultaneously and we use $\mathcal{NR}$ to trace them.
In line 7, we start to verify the essential tuples of $e$ from the one with the highest overlapping degree. 
For each essential tuple $\{u,e,s_u\}$, i.e., $(u, s_u)$ in $\mathcal{D}(e)$, the supporting check for reachability $u {\leadsto} e {\leadsto} v$ with all unprocessed $v$ is performed in lines 9-13.
$\{u,e,s_u\}$ is determined not redundant if $\mathcal{S}$ is missing some unprocessed vertices or it has been marked as not redundant during the verification of a previous essential tuple of $e$.
If $\{u,e,s_u\}$ is not redundant, line 15 adds it to $\mathcal{L}^*$ and line 16 marks $\{w,e,s_u\}$ as not redundant for those unprocessed vertices $w$ not contained in $\mathcal{S}$; otherwise, we remove $\{u,e,s_u\}$ from $\mathcal{L}$ in line 18. $(u, s)$ is removed from $\mathcal{D}(e)$ in line 19 to reduce further computation since all unprocessed vertices $v \in \mathcal{D}(e)$ s.t. $u \leadsto e \leadsto v$ and only supported by $e$ have been recorded in $\mathcal{NR}$ set. 
Line 20 terminates the verification process when all unverified essential tuples of $e$ are already marked as not redundant, and we add them to $\mathcal{L}^*$ in line 21. 
Given $l$ be the number of entries in $\mathcal{L}$. Since $\mathcal{L}$ and $\mathcal{D}$ are dual structures, we have $\sum^m_{i = 1} \left|\mathcal{D}(i)\right| = l$, which indicates that line 4 in Algorithm \ref{alg:minimalnew} will be called at most $l$ times. We can then analyze the complexity of Algorithm \ref{alg:minimalnew} as follows.


\begin{theorem} [Time Complexity]
\label{theo:mini:timecomp}
    
    The time complexity of Algorithm~\ref{alg:minimalnew} is $O(l \cdot \log\theta \cdot( l_v \beta  + \theta))$, 
    where $l_v = \max_{u \in \mathcal{V}}{\left|\mathcal{L}(u)\right|}$, $\theta = \max_{e \in \mathcal{E}}{\left|\mathcal{D}(e)\right|}$, and $\beta = \max_{e \in \mathcal{E}}{\left|\mathcal{I}(e)\right|}$, where $\beta \le \theta$.
\end{theorem}


\begin{theorem} [Minimality Analysis]
\label{theo:minianalyz}
 The HL-index $\mathcal{L}^*$ generated by Algorithm~\ref{alg:minimalnew} for a hypergraph $\mathcal{H} = (\mathcal{V, E})$ is minimal, which satisfies the following two properties: $i)$ \textbf{Completeness.} The $MR(u, v)$ query for any pair of vertices $u,v \in \mathcal{V}$ can be correctly answered by $\mathcal{L}^*$, and $ii)$ \textbf{Necessity.} For any vertex $u$ and an arbitrary label in $\mathcal{L}^*(u)$, at least one $MR(u, *)$ query will be incorrectly answered if we remove $(e, s)$ from $\mathcal{L}^*(u)$.
    
\end{theorem}

\noindent \textbf{HL-index Maintenance.} 
We further study maintenance of the HL-index under hyperedge updates.
Vertex insertions and deletions can naturally reduce to sequences of hyperedge operations.
When a new hyperedge is inserted, it may introduce additional walks that expand reachability between vertices, while simultaneously rendering some existing entries redundant. In this case, pruning strategies are required to preserve the minimality of the HL-index. In contrast, hyperedge deletions may invalidate entries that rely on the removed hyperedge, and when no alternative walks exist to sustain the same reachability, the corresponding entries must be removed. Moreover, deletions may also trigger the recovery of previously pruned entries once their dominating alternatives disappear, requiring iterative verification until the index stabilizes.

\section{Query Processing} \label{sec:5}

\begin{algorithm}[t]
	{
		\SetVline
		\footnotesize
		\caption{{\textsc{VTE}-reach Query Algorithm}}
		\label{alg:alg6}
		\Input{An HL-index $\mathcal{L}$, source vertex $u$, target vertex $v$}
		\Output{The result of $MR(u, v)$}
		\State  {$k \leftarrow 0, i \leftarrow 1, i' \leftarrow 1$}
    
        \While{$i \leq \left|\mathcal{L}(u)\right| \land i' \leq \left|\mathcal{L}(v)\right|$} {
            \State {$(e, s_u) \leftarrow$ $i$-th element in $\mathcal{L}(u)$}
            \State {$(e', s_v) \leftarrow$ $i'$-th element in $\mathcal{L}(v)$}
            \State{\textbf{if}~$s_u \leq k \lor \mathcal{O}(e) < \mathcal{O}(e')$~\textbf{then}~$i \leftarrow i + 1$}
            \State{\textbf{else if}~$s_v \leq k \lor \mathcal{O}(e) > \mathcal{O}(e')$~\textbf{then}~$i' \leftarrow i' + 1$}
            \lElse {
                \State {$k \leftarrow \min(s_u, s_v)$, $i \leftarrow i + 1$, $i' \leftarrow i' + 1$}
            }
        }
        \Return{$k$}
	}
\end{algorithm}

In this section, we present the query processing approaches for the max-reachability and $s$-reachability problem using the HL-index, whose details are based on the following lemma.


\begin{lemma}
    \label{lem:mrquery}
    Given the HL-index $\mathcal{L}$ of a hypergraph $\mathcal{H} = (\mathcal{V, E})$, and $u, v \in \mathcal{V}$, 
    $MR(u, v) = \max\{\min(s_u, s_v) | (e_u, s_u) \in \mathcal{L}(u) \land \ (e_v, s_v) \in \mathcal{L}(v) \land e_u = e_v\}$.
\end{lemma}



\noindent\textbf{Max-reachability query processing.} 
According to Lemma~\ref{lem:mrquery}, $MR(u,v)$ can be answered by finding hyperedges that appear in both $\mathcal{L}(u)$ and $\mathcal{L}(v)$.
Since entries in the HL-index for each vertex are in descending order based on the target hyperedge, finding hyperedge in both $\mathcal{L}(u)$ and $\mathcal{L}(v)$ can be achieved in a merge-sort manner.
Once we find a hyperedge $e$ s.t. $(e, s_1) \in \mathcal{L}(u) \land (e, s_2) \in \mathcal{L}(v)$, we update result to $\min(s_1, s_2)$. We keep scanning the $\mathcal{L}(u)$ and $\mathcal{L}(v)$ until either of them reaches its end. The pseudocode is shown in Algorithm \ref{alg:alg6}.

\begin{theorem} [Time Complexity] \label{th:queryComplexity}
    Given $u,v \in \mathcal{V}$, the time complexity of Algorithm \ref{alg:alg6} is bounded by $O(\left|\mathcal{L}(u)\right| + \left|\mathcal{L}(v)\right|)$.
\end{theorem}




\noindent$\boldsymbol{s}$\textbf{-reachability query processing.} Given two query vertices $u$, $v$ and an integer $s$, we have $u \stackrel{s}{\leadsto} v$ iff $MR(u,v) \ge s$.
Therefore, the $s$-reachability query can be derived from a modified version of Algorithm \ref{alg:alg6}. To answer if $u \stackrel{s}{\leadsto} v$, we first initialize the $k$ in Algorithm \ref{alg:alg6} to $s - 1$ at line 1. We can return $true$ if it reaches the else-branch at line 7.





\section{Experiment} \label{sec:6}



In this section, we evaluate the performance of our proposed techniques by experimenting on 20 hypergraphs.

\subsection{Experiment Setup}

\myparagraphexp{Algorithms}
We summarize the algorithms used as follows: 
\begin{itemize}[leftmargin=*]
\item HypED\cite{preti2024hyper}: approximate method for $s$-walk distance estimation, we have modified it to adapt to our
problem.
\item Base: Bidirectional online search (Algorithm~\ref{alg:BASELINE}).
\item Base*: Base integrated with precomputed neighbors.
\item Base-Construct: Basic HL-index construction (Algorithm~\ref{alg:alg3}). 
\item Construct: Fast HL-index construction (Algorithm~\ref{alg:alg4}). 
\item Construct*: Minimal HL-index construction method that extends Algorithm \ref{alg:alg4} by combining the minimal index generating technique mentioned in Algorithm~\ref{alg:minimalnew}.

\item VTE-reach: Algorithm~\ref{alg:alg6} on index from Construct.
\item Min-reach: Algorithm~\ref{alg:alg6} on index from Construct*.
\item ETE-reach: Described in Section~\ref{th:sec3.2}, ETE-index-based, reduces computation cost via a merge-sort-like strategy.
\end{itemize}

\begin{table}[t]

\caption{Statistics of datasets}
\label{tab:tab1}
\footnotesize
\centering
\resizebox{0.98\linewidth}{!}{%
\begin{tabular}{lcccc}
\toprule

\textbf{Dataset} & \textbf{$\left|\mathcal{V}\right|$} & \textbf{$\left|\mathcal{E}\right|$} & \textbf{$\eta_{avg}$} & \textbf{$\eta_{max}$}   \\

\midrule
NC (NDC-classes) & 1,161 & 1,222 & 5 & 221\\
NS (NDC-substances) & 5,311 & 10,025 & 9 & 578 \\
SS (small-world) & 10,000 & 10,000 & 6.6 & 55 \\
BK (BrightKite) & 4,301 & 5,201 & 3.9 & 54\\

PS (primary-school) & 242 & 12,704 & 126 & 261 \\ 
PE (phs-email) & 4,423 & 15,653 & 14 & 4,869\\
BR (cat-Brain) & 638 & 37,250 & 116 & 292  \\
EE (email-Eu) & 998 & 25,791 & 85 & 910  \\
CK (cat-Cooking) & 6,714 & 39,774 & 63 & 18,049 \\
HB (house-bills) & 1,494 & 60,987 & 835 & 6,220\\
WA (walmart-trips) & 88,860 & 69,906 & 5 & 5,733\\
CD (cat-DAWN) & 2,109 & 87,104 & 162 & 16,090 \\
DA (DAWN-unique) & 2,558 & 143,523 & 216 & 25,876 \\
TU (tags-ask-ubuntu) & 3,029 & 151,441 & 164 & 12,930 \\
TS (tags-math) & 1,629 & 174,933 & 160 & 40,682 \\
TM (threads-math) & 176,444 & 595,747 & 7 & 11,358 \\
DB (coauth-DBLP) & 1,930,378 & 3,700,681 & 4 & 1,399 \\
AM (amazon-reviews) & 2,268,231 & 4,285,363 & 32 & 28,973\\
SO (stackoverflow) & 2,662,150 & 6,964,551 & 5 & 194 \\
LJ (livejournal) & 3,201,203 & 7,489,073 & 33 & 289 \\
\bottomrule
\end{tabular}
}
\end{table}

\myparagraphexp{Datasets}
We employ 17 real-world and 3 synthetic hypergraphs (SS, BK, LJ) in our experiments, with details shown in Table \ref{tab:tab1}. SS is generated by HGX \cite{lotito2023hypergraphx}.  
All other datasets are obtained or synthesized from cornell\footnote{\url{https://www.cs.cornell.edu/~arb/data/}}, {konect}\footnote{\url{http://konect.cc/networks/}}, {SNAP}\footnote{\url{https://snap.stanford.edu/data/}} and \cite{preti2024hyper}, which are publicly available. 
We propose and apply several graph compacting techniques for all datasets to reduce redundant information.
Due to the space limitation, we present these strategies in Appendix B online~\cite{appendix}. 

\myparagraphexp{Settings and workloads} Queries that cannot finish within 10 hours or out of memory are set as \textbf{INF}.
In the experiment, all algorithms are implemented in standard C++ and compiled with g++ 11.4.0 with -O3 optimization. All experiments are performed on a machine with Intel Xeon 2.80GHz and 503GB main memory running Linux (Ubuntu 22.04.1 LTS, 64-bit).



    

    

\subsection{Experimental Results}

\begin{figure} [t]
    \centering
    \includegraphics[width=0.82\linewidth]{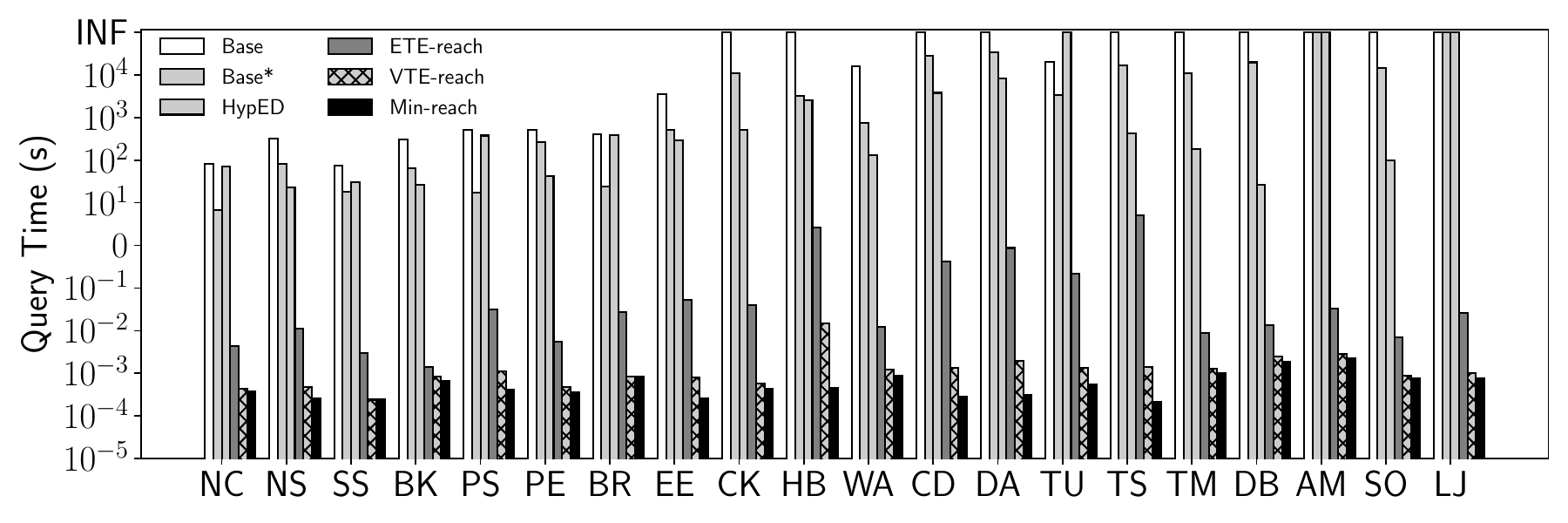}
     \caption{Total query time on all the datasets}
    \label{fig:fig5}
\end{figure}

\myparagraphexp{Exp-1: Query time on all the datasets} 
\label{sec:QueryTime}
In this experiment, we randomly pick 1,000 pairs of vertices and report the aggregated $MR$ query time of all mentioned query methods on each dataset, as shown in Figure~\ref{fig:fig5}.
As observed, Base* runs faster than Base among all datasets, =while index-based solutions achieve a speedup of at least 3 orders of magnitude compared with online solutions in most datasets. HypED can finish the 1,000 queries in tens of seconds for several datasets, but it will incur out of memory issue (e.g., when building oracles for TU, AM, LJ), since the maximum value of $s$ can be equal to $\delta$ when querying for max-reachability. We also notice that Min-reach significantly outperforms all others, with over 4 orders of magnitude speedup compared with Base*, at least 2 orders of magnitude for the ETE-reach, HyPED and faster than VTE-reach among all the datasets. For example, on {TS}, Base cannot finish in a reasonable time. 
Base*, ETE-reach and VTE-reach return the result in 4 hours, 5.2 seconds and 1.4ms, respectively, while for our minimal index-based algorithm Min-reach, all queries are processed within 0.3ms. 
Additional results under varying query vertices degree are provided in Appendix D \cite{appendix} due to limited space.

\myparagraphexp{Exp-2: Indexing time}
We report the indexing time of Construct-Base, Construct and Construct* on all datasets. 
For Construct-Base, we implement line 8 of Algorithm~\ref{alg:alg3}
by first filtering out the cases where $e_w$ cannot exist. Then we conduct a bidirectional BFS search to verify the existence of covering relationships. We denote the Indexing time of Construct-Base, Construct and Construct* as $\mathcal{C}$, $\mathcal{T}, \mathcal{T}^*$, respectively.
For indexing time over 10 hours, we set it as INF.
The results are shown in Table \ref{tab:tab2}. It is clear that both Construct and Construct* perform orders of magnitudes faster than the Construct-Base, while Construct-Base exceeds the time limit for many datasets. We also notice that the additional cost of minimal HL-index generation is slight compared with the construction cost of computing a complete HL-index. For example, for DB, it takes 1,950 seconds to build a complete HL-index, while extending it to a minimal HL-index only affords an extra 39 seconds.

\setlength{\tabcolsep}{4pt}
\begin{table}[t]
\renewcommand{\arraystretch}{1}
\caption{{Construction cost on all the datasets}}
\label{tab:tab2}
\centering
\resizebox{\linewidth}{!}{
\begin{tabular}{cccc|ccccc}
\toprule
\multirow{2}{*}{\textbf{Dataset}} & \multicolumn{3}{c|}{\textbf{Construction Time}} & \multicolumn{5}{c}{\textbf{Space Cost}} \\
 & \multicolumn{1}{c}{\textbf{$\mathcal{C}$}} & \textbf{$\mathcal{T}$} & \textbf{$\mathcal{T}^*$} & \textbf{$\mathcal{H}$} & \textbf{$\mathcal{L}$} & \multicolumn{1}{c}{\textbf{$\mathcal{L}^*$}} & \textbf{$\mathcal{N}$} & \textbf{$\mathcal{\hat{M}}$} \\ \hline
\textbf{NC} & 31s & 0.1s & 0.11s & 25KB & 28KB & 24KB & 765KB & 41KB \\
\textbf{NS} & 9,204s & 1.2s & 1.78s & 194KB & 378KB & 260KB & 52MB & 292KB \\
\textbf{SS} & 4,801s & 0.15s & 0.23s & 220KB & 554KB & 553KB & 7MB & 18KB \\

\textbf{BK} & 2,719s & 0.3s & 0.43s & 64KB & 130KB & 121KB & 10MB & 129KB \\

\textbf{PS} & 6,389s & 0.9s & 1.03s & 120KB & 164KB & 49KB & 50MB & 19KB \\
\textbf{PE} & 29,058s & 11s & 11.3s & 251KB & 204KB & 163KB & 457MB & 826KB \\
\textbf{BR} & 5,516s & 3.2s & 4.3s & 291KB & 461KB & 461KB & 119MB & 126KB \\
\textbf{EE} & 14,329s & 4.5s & 4.64s & 332KB & 257KB & 75KB & 183MB & 437KB \\
\textbf{CK} & INF & 320s & 390s & 1.7MB & 3.3MB & 1.7MB & 7.3GB & 65KB \\
\textbf{HB} & INF & 438s & 453s & 4.8MB & 7.5MB & 0.41MB & 786MB & 124KB \\
\textbf{WA} & INF & 22.7s & 33s & 1.8MB & 3.79MB & 3.75MB & 695MB & 596KB \\
\textbf{CD} & INF & 303s & 319s & 1.4MB & 1.8MB & 0.35MB & 9.8GB & 79KB \\
\textbf{DA} & INF & 1,020s & 1,055s & 2.1MB & 2.7MB & 0.47MB & 27GB & 166KB \\
\textbf{TU} & INF & 302s & 334s & 1.0MB & 2.8MB & 0.69MB & 12GB & 104KB \\
\textbf{TS} & INF & 593s & 687s & 1.0MB & 1.8MB & 0.06MB & 23GB & 324KB \\
\textbf{TM} & INF & 379s & 525s & 5.6MB & 11MB & 8.1MB & 14GB & 342KB \\
\textbf{DB} & INF & 1,950s & 1,989s & 37MB & 60MB & 55MB & 52GB & 61MB \\
\textbf{AM} & INF & 19,700s & 23,608s & 280MB & 577MB & 440MB & 295GB & 1.0MB \\
\textbf{SO} & INF & 6,998s & 7,065s & 47MB& 108MB & 107MB & 5.6G& 2.1MB \\
\textbf{LJ} & INF & 21,653s & 24,312s & 396MB & 811MB& 634MB & 2.1GB& 2.0MB \\
\bottomrule

\end{tabular}
}

\end{table}

\myparagraphexp{Exp-3: Space cost of index construction}
In Table~\ref{tab:tab2}, we report the
size of the original graph (i.e., $\mathcal{H}$), 
the proposed HL-index built by Construct and Construct* (i.e., $\mathcal{L}$ and $\mathcal{L}^*$, respectively) in all datasets. 
As observed,
an HL-index $\mathcal{L}$ incurs a much larger space cost than its corresponding minimal index $\mathcal{L}^*$ in most datasets, which also implies a longer query time on $\mathcal{L}$. 
For example, in TS, $\mathcal{L}$ consumes 1.8MB, which is 30x larger than the space cost of $\mathcal{L}^*$. 
Moreover, the size of our minimal HL-index is comparable to the original graph, even smaller in some datasets.
For example, in the dataset {EE}, $\mathcal{L}^*$ occupies 75KB, while the original graph size is 332KB.

\begin{figure}[t] 
  \centering
  \begin{subfigure}[b]{0.32\columnwidth} 
    \centering
    \includegraphics[width=\linewidth]{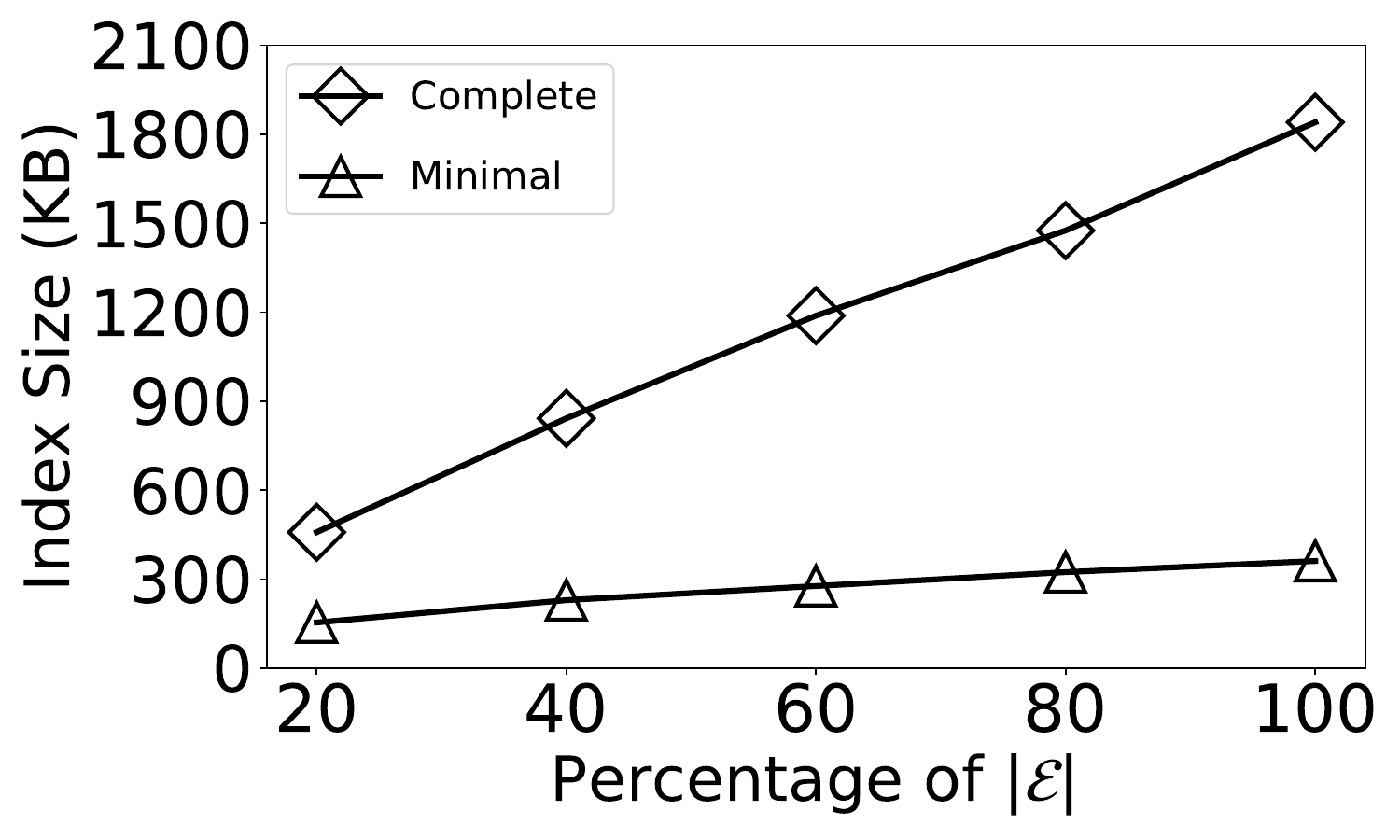}
    \caption{CD}
  \end{subfigure}
    \hfill
    \begin{subfigure}[b]{0.32\columnwidth} 
    \centering
    \includegraphics[width=\linewidth]{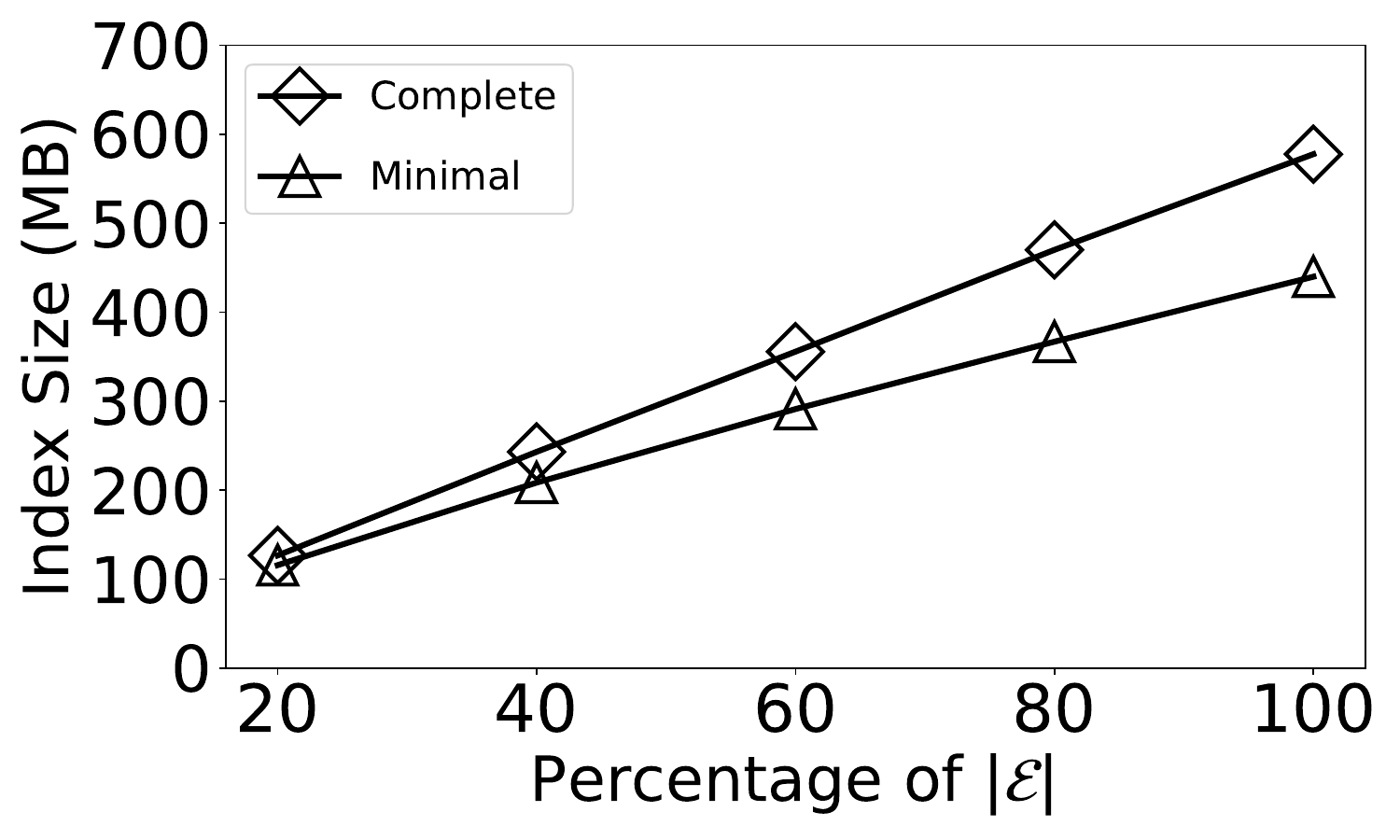}
    \caption{AM}
  \end{subfigure}
  \hfill
  \begin{subfigure}[b]{0.32\columnwidth} 
    \centering
    \includegraphics[width=\linewidth]{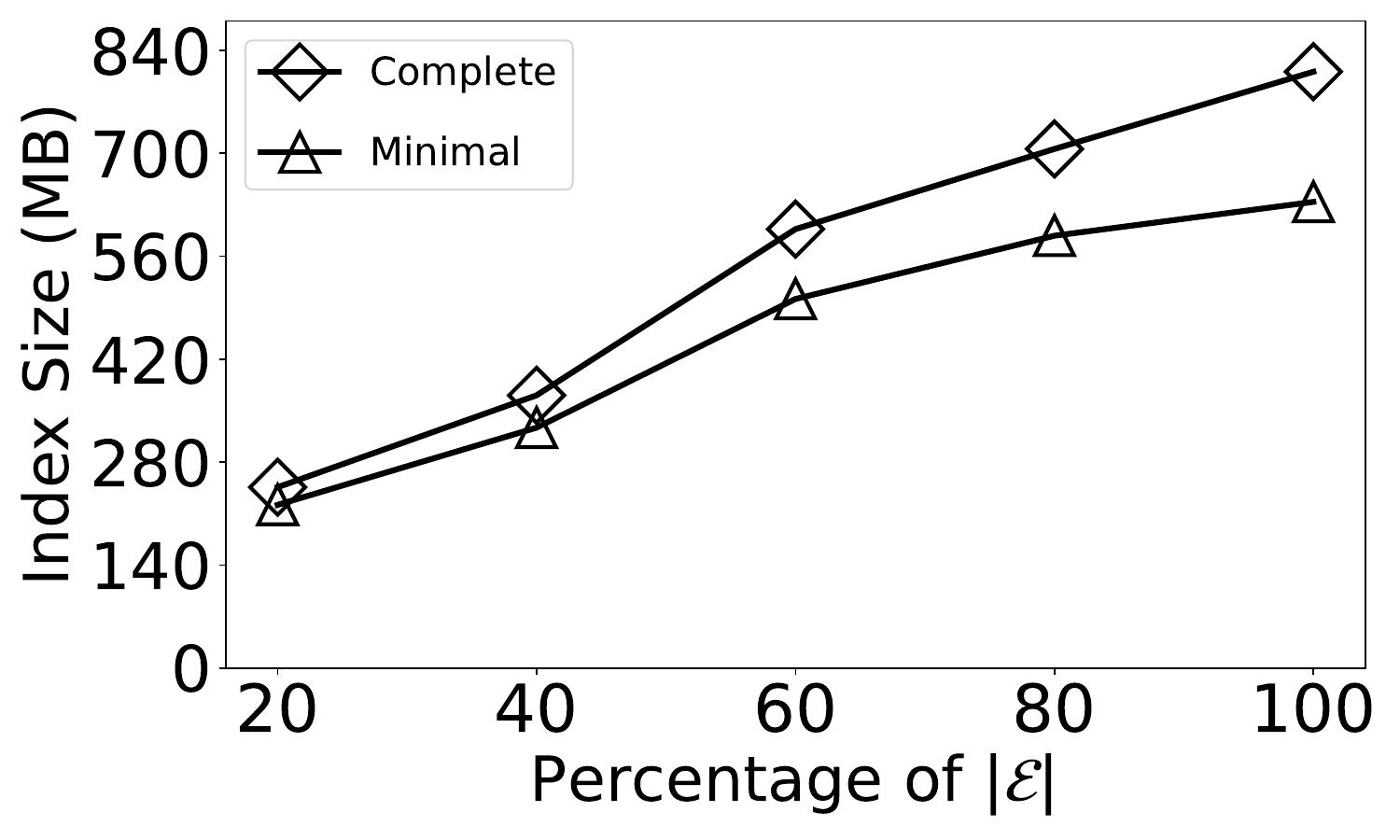}
    \caption{LJ}
    
    \label{fig:svg1}
  \end{subfigure}

    \vspace{1mm}
  \hfill
  \begin{subfigure}[b]{0.32\columnwidth} 
    \centering
    \includegraphics[width=\linewidth]{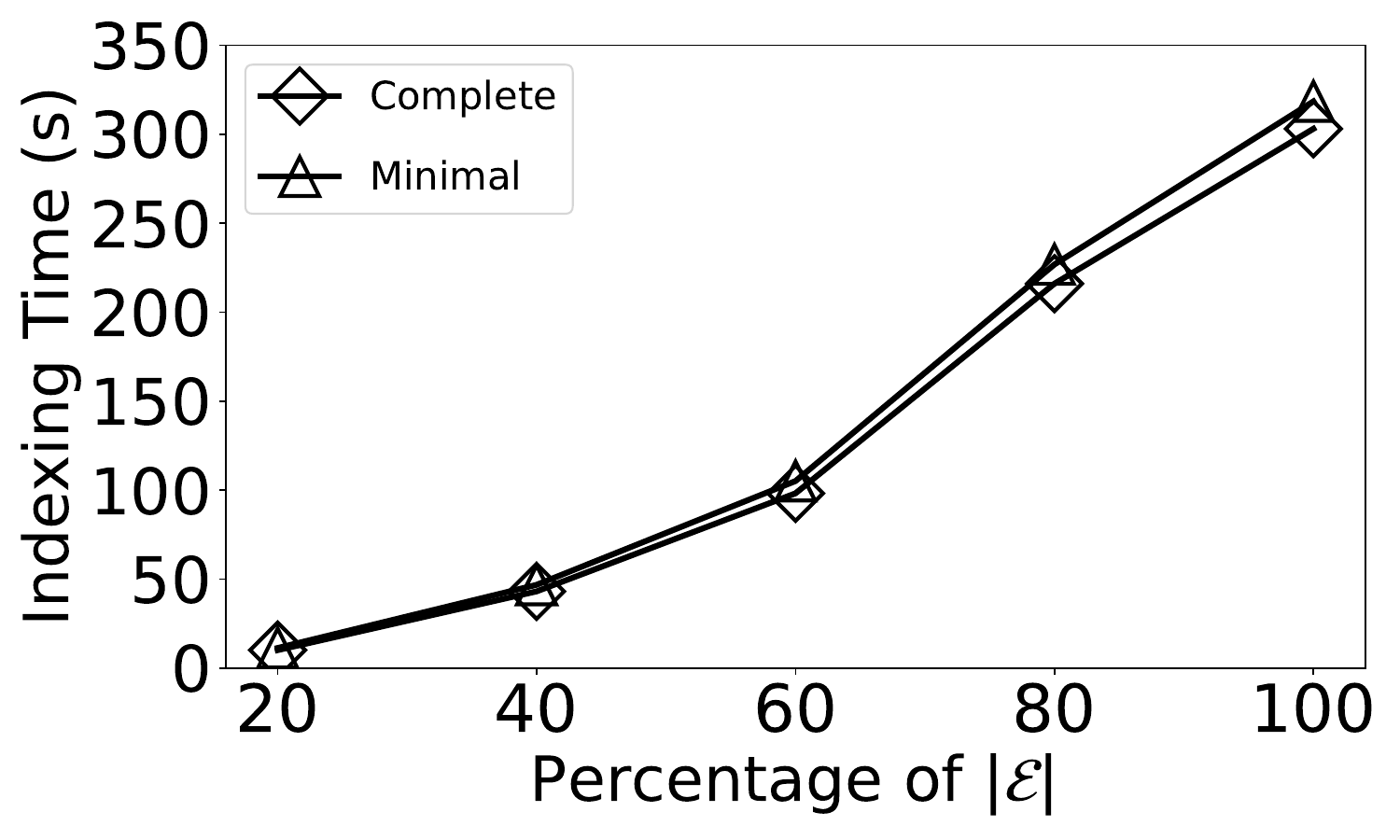}
    \caption{CD}
  \end{subfigure}  
  \hfill
  \begin{subfigure}[b]{0.32\columnwidth} 
    \centering
    \includegraphics[width=\linewidth]{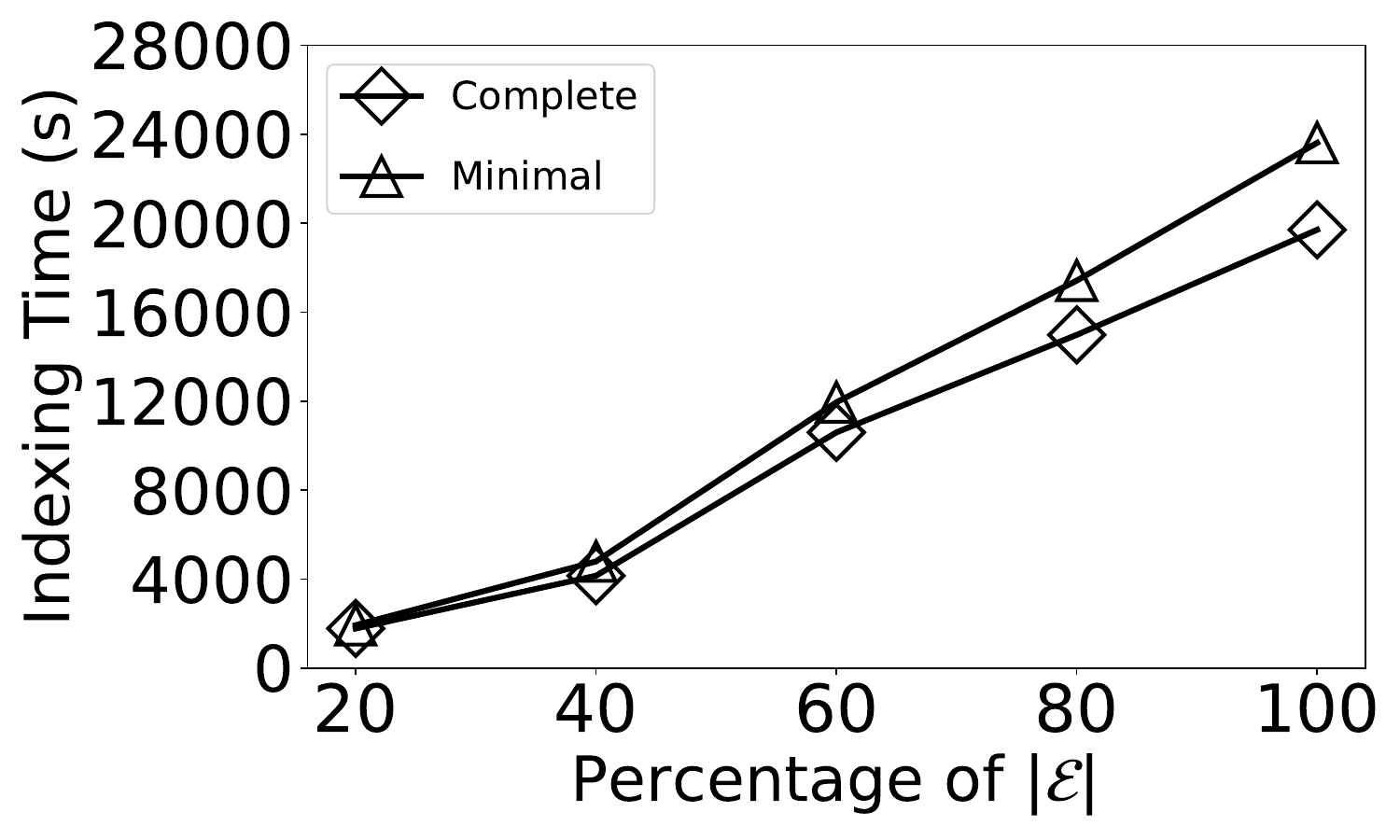}
    \caption{AM}
  \end{subfigure}
  \hfill
  \begin{subfigure}[b]{0.32\columnwidth} 
    \centering
    \includegraphics[width=\linewidth]{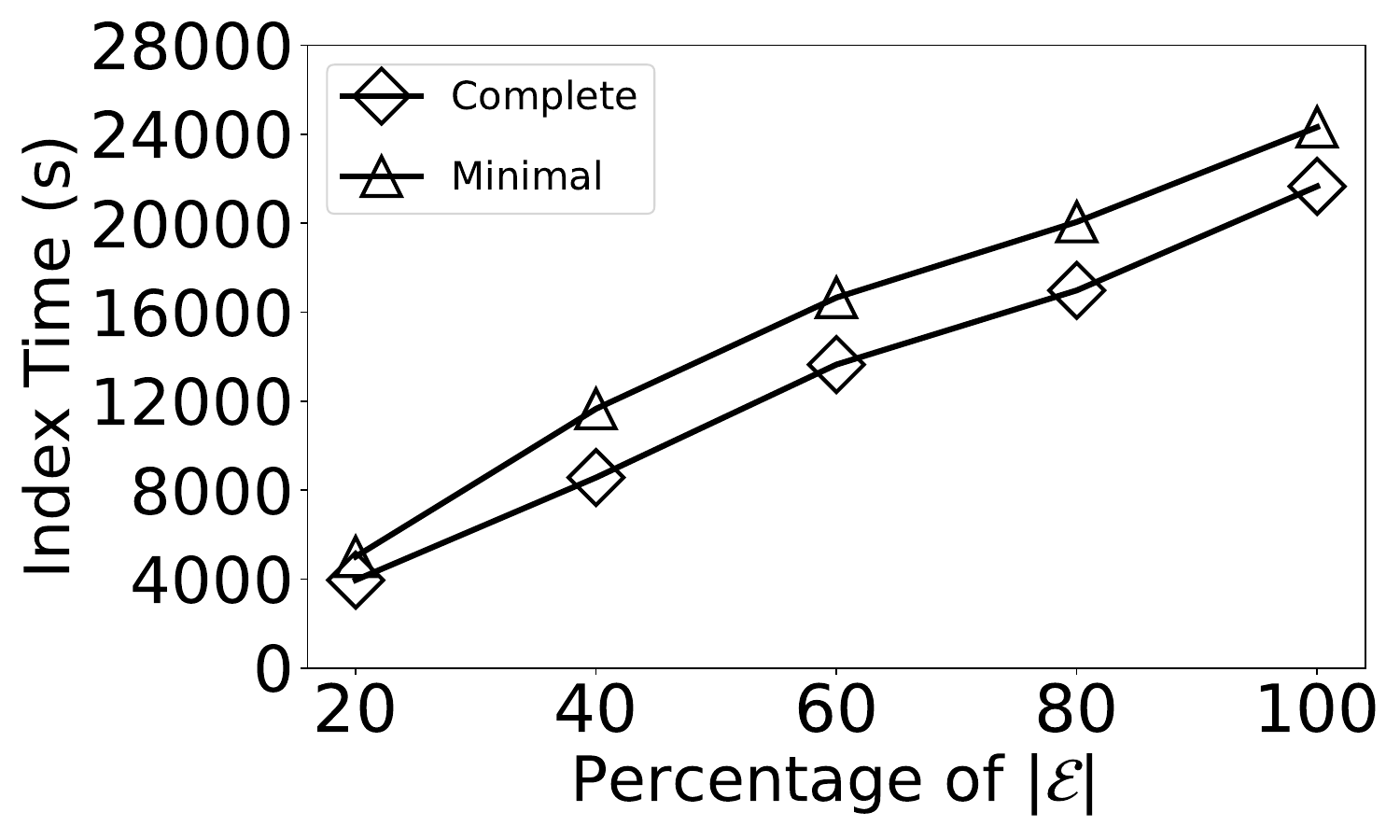}
    \caption{LJ}
    
    \label{fig:svg3}
  \end{subfigure}

  \caption{Scalability evaluation of index construction}
  \label{fig:fig7}
\end{figure}

Next, we demonstrate the neighbor computation optimization proposed in Section~\ref{sec:5.3}.
Naively, all neighbor information should be maintained in an adjacency list, its size is denoted as $\mathcal{N}$, as reported in Table \ref{tab:tab2}.
Alternatively, we use $\hat{\mathcal{M}}$ to represent the peak size of neighbor-index $\mathcal{M}$, where $\mathcal{M}$ is initialized as empty and updated during index construction. 
As shown in Table~\ref{tab:tab2}, the size of $\hat{\mathcal{M}}$ is much smaller than $\mathcal{N}$. By our hyperedge ordering strategy and Lemma~\ref{th:lemma10}, $MCD(e)$ exceeds the $s$ values for many of $(e',s)\in\mathcal{N}(e)$, so only a small fraction of neighbors are kept in $\mathcal{M}(e)$, greatly reducing space cost. Our neighbor-index occupies memory close to the original graph size yet is orders of magnitude smaller than $\mathcal{N}$. For example, in dataset DB it requires only 61MB (vs. 37MB graph size), compared with over 52GB for $\mathcal{N}$. In dataset AM, it uses merely 1MB, while $\mathcal{N}$ consumes 300GB, since high-importance hyperedges with large overlaps allow many $MCD$ values to be resolved early, leaving few neighbors to store.

\myparagraphexp{Exp-4: Scalability evaluation of index construction}
We evaluate the scalability of both Construct and Construct* in 3 datasets, {CD}, {AM} and {LJ}.
Specifically, for each dataset, we randomly select 20\%-100\% hyperedges from $\mathcal{E}$ to form 4 new subgraphs. 
Figure \ref{fig:fig7}(a)-\ref{fig:fig7}(c) (resp. Figure \ref{fig:fig7}(d)-\ref{fig:fig7}(f)) report the index size (resp. indexing time) of the HL-index in these datasets.
As observed, the indexing time for both the complete and minimal HL-index grows approximately linearly with the increase in graph size. Furthermore, the index size of a complete HL-index increases in an approximately linear trend, while the uptrend is flatter for the minimal HL-index, demonstrating the scalability of our proposed methods.
For example, as shown in Figure \ref{fig:fig7}(b) and \ref{fig:fig7}(e), when the percentage of $|\mathcal{E}|$ is 40\% in AM, the index size and indexing time for both the complete and minimal HL-index are 243MB, 208MB, 4,152s and 4,794s, respectively.
When the percentage of $|\mathcal{E}|$ is 100\%, the index size and construction time are 577MB, 440MB, 19,700s and 23,608s, respectively.

\begin{figure} [t]
    \centering
    \includegraphics[width=0.9\linewidth]{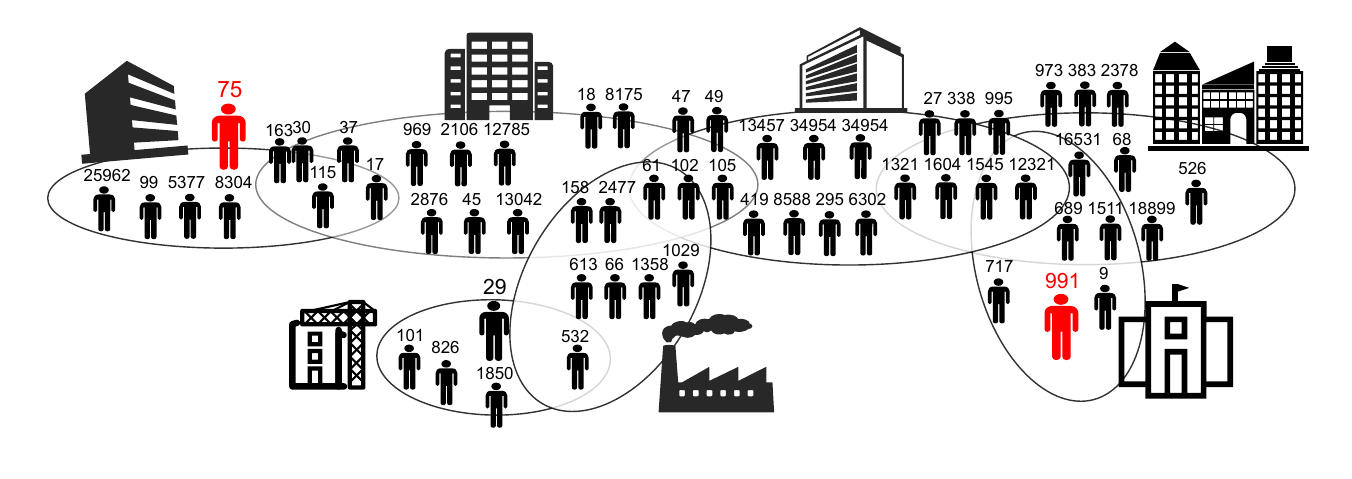}

     \caption{Case study for epidemic network}
    \label{fig:case1}
\end{figure}

\myparagraphexp{Exp-5: Case study}
To demonstrate the effectiveness of our proposed max-reachability model, we conduct a case study on the dataset Brightkite by constructing a hypergraph across a 21-day period (6th-26th March 2009), consistent with typical incubation cycles, to capture the potential transmissions.
Users who check in at the same location on the same day are grouped into a hyperedge. 
Given an infected individual, a natural question is whether another person is also infected or at risk of being affected, and to what extent. To answer this, we can compute the max-reachability between the two individuals, which quantifies the strength of their potential connection. For example, given a query on individuals 75 and 991, we illustrate in Figure~\ref{fig:case1} the partial path corresponding to their max-reachability value.
In this figure, each person is labeled by their ID, and each ellipse refers to a hyperedge, representing users who checked in at the same location on the same day. The max-reachability between individuals 75 and 991 is 5, suggesting that they may be connected through a latent transmission chain. Intuitively, a larger max-reachability value indicates a stronger potential connection between individuals, since repeated co-locations at the same places naturally increase the likelihood of exposure and infection. 
In a nutshell, this case study highlights the practical utility of max-reachability in analyzing contagion pathways and quantifying individual risks, providing additional perspectives for epidemic analysis.

\section{Related Work} \label{sec:7}


\noindent \textbf{Reachability query in pairwise graphs.} {
    Graphs are widely used to model
relationships between entities across various domains \cite{zai2025proh,tan2025paths, tan2025hydraRAG,tan2025memotime,li2025efficient,DBLP:conf/icde/WuSWZ00024,DBLP:journals/pvldb/WuSWWZQL24,DBLP:journals/pvldb/WangWWZQZL24,DBLP:conf/www/WangWWCZ025,wu2025efficient}. Extensive studies have been conducted on reachability queries in pairwise graphs \cite{agrawal1989efficient, chen2008efficient, cheng2013tf, cohen2003reachability, wei2014reachability, zhu2014reachability,  su2016reachability, yuan2024efficient}, 
    and the 2-hop labeling framework is widely adopted to design efficient index-based solutions~\cite{cohen2003reachability, jin20093, zhu2014reachability, chvatal1979greedy, schenkel2004hopi}.
     Cohen et al. first introduced the concept of the 2-hop labeling framework in \cite{cohen2003reachability}.
     TOL \cite{zhu2014reachability} is a general total order approach for computing the optimal 2-hop index by assigning a unique rank for each vertex in the graph, while TFL \cite{yano2013fast} and PLL \cite{zhu2014reachability} adopt the idea of TOL by starting an initialisation of the total order based on topological and vertex degree, respectively. Schenkel et al. \cite{schenkel2004hopi} reduce the complexity of building the 2-hop index construction from $O(n^4)$ to $O(n^3)$. 
     A large number of works have also been done to maintain the 2-hop index in dynamic pairwise graphs, including DBL \cite{lyu2021dbl}, U2-hop \cite{bramandia2009incremental} and HOPI~\cite{schenkel2005efficient}.  Interested readers can find more details in the survey~\cite{zhang2023overview}. Moreover, the 2-hop labeling framework is also widely used to support different types of graphs, such as temporal graphs \cite{wen2022span, semertzidis2015timereach} and bipartite graphs \cite{chen2021efficiently}. However, these techniques cannot be trivially extended to support the case in hypergraphs. 

}

\myparagraph{Hypergraph analysis} Hypergraph structure has been employed to model high-order correlation among data in recent years from different scopes, such as graph clustering \cite{takai2020hypergraph, leordeanu2012efficient, li2024hypergraph}, dense subgraph mining \cite{luo2021hypercore} and community search  \cite{zhang2022hyperiso,DBLP:journals/pvldb/ArafatKRG23}. 
Some studies have been conducted about path-related problems in hypergraphs.  Gallo et al. \cite{gallo1993directed} study the problem of finding the minimum cost $B$-path in weighted hypergraphs, and Gao et al. \cite{DBLP:journals/ton/GaoZRSRB15} introduce the concept of the shortest path in both static and dynamic hypergraphs. 
Shun et al. \cite{shun2020practical} explore the way to achieve high-performance hyperpath processing through efficient parallel algorithms. 
Liu et al. \cite{liu2024reordering} extend the concept of pairwise graph ordering \cite{wei2016speedup} to hypergraphs and then propose the compression-array acceleration
structure to improve memory access efficiency and further speed up the traversal in hypergraphs. 
HypED \cite{preti2024hyper} introduces a framework for estimating $s$-distance, which is unsuitable for our problem because:
(i) it only supports queries with a fixed $s$ (set to 10 in their paper), whereas max-reachability requires varying $s$ up to tens of thousands depending on hyperedge sizes, incurring excessive overhead; and
(ii) it maintains information only between hyperedges, making it inefficient for vertex-to-vertex queries. Thus, HypED is not suitable for our problem.
\section{Conclusion} \label{sec:8}

To effectively capture the groupwise reachability relationships among entities in hypergraphs,
we introduce a novel $s$-reachability model in this paper. 
Moreover, 
we further propose and investigate the max-reachability problem, which is a generalized version of the $s$-reachability model. 
To scale for large graphs,
we design a robust index-based framework called HL-index, supported by several optimization techniques that improve both the time and space efficiency of index construction. 
We first design an efficient approach to detect the covering relationship in hypergraphs. Then, a lightweight neighbor-index is designed to accelerate the index construction, while affording a slight space cost. 
We prove that our HL-index holds the properties of both correctness and minimality. Extensive experiments on 20 datasets are conducted to validate the performance of our proposed method. 

\noindent \textbf{Acknowledgments}.
This work is partially supported by
ARC DP240101322, DP230101445, FT210100303.
Yanping Wu and Xiaoyang Wang are the corresponding authors.

    

\section*{AI-Generated Content Acknowledgment} \label{sec:ack}

The authors used ChatGPT only for proofreading and small language refinements. All technical content, experimental design, and analysis were done entirely by the authors without help from ChatGPT or other AI tools.
\bibliographystyle{IEEEtran}
\bibliography{main}


\end{document}